\documentclass[epj]{svjour}
\usepackage[T1]{fontenc}
\usepackage[utf8]{inputenc}
\usepackage[british]{babel}
\usepackage{graphics}
\usepackage{graphicx}
\usepackage{longtable}
\usepackage{color}
\usepackage{sverb}
\usepackage{cite}
\usepackage{array,booktabs}
\usepackage{placeins}

\usepackage[caption=false]{subfig}
\usepackage[nointegrals]{wasysym}
\usepackage{amsmath}
\usepackage[htt]{hyphenat}
\usepackage{listings}
\definecolor{airforceblue}{rgb}{0.36, 0.54, 0.66} 
\definecolor{amaranth}{rgb}{0.9, 0.17, 0.31}
\usepackage{url}
\begin{document}
\sloppy


\title{
	Production of $\tau\tau jj$ final states at the LHC and  the  {\texttt{TauSpinner}} algorithm: the spin-2 case
}
\author{
	M. Bahmani\inst{1} \and J. Kalinowski\inst{2} \and W. Kotlarski\inst{2,3} \and E. Richter-W\k{a}s\inst{4} \and Z. W\k{a}s\inst{1,}\thanks{\emph{e-mail:} Z.Was@cern.ch}
}
\institute{Institute of Nuclear Physics, PAN, Krak\'ow, ul. Radzikowskiego 152, Poland 
	\and Faculty of Physics, University of Warsaw, Pasteura 5, 02-093 Warsaw,  Poland
    \and IKTP, Technische Universit\"at Dresden, Zellescher Weg 19, 01069 Dresden, Germany
	\and Institute of Physics, Jagellonian University, Lojasiewicza 11, 30-348 Cracow, Poland
}
\abstract{
The {\tt TauSpinner} algorithm is a tool that allows to 
modify the physics model 
of the Monte Carlo generated samples due to the changed assumptions of event production dynamics, 
but without the need of re-generating events. With the help of weights  $\tau$-lepton production 
or decay processes can be modified accordingly to a new physics model.
 In a recent paper a new version 
{\tt TauSpinner ver.2.0.0} has been presented which includes a provision for  introducing non-standard  states and couplings 
and study their effects in the vector-boson-fusion processes by exploiting  the spin 
correlations of $\tau$-lepton pair 
decay products in processes where final states include also two hard jets. In the present paper we  document how this can be achieved taking 
as an example the non-standard spin-2  state that couples to Standard Model particles and 
 tree-level matrix elements with complete helicity information   included  for the parton-parton scattering amplitudes into 
a $\tau$-lepton pair and two outgoing partons.
This implementation is prepared as the external (user provided) routine for the {\tt TauSpinner} algorithm.
It exploits amplitudes 
generated by {\tt MadGraph5}  and adopted to the {\tt TauSpinner} algorithm format. 
Consistency tests of the implemented matrix elements, reweighting algorithm and numerical results 
{  for observables sensitive to $\tau$ polarization} are presented. 
}
\maketitle

\section{Introduction}

With increasing statistics collected by the  LHC experiments the interests to explore final states 
with $\tau$-leptons gain importance. Because of the high mass and their decays, $\tau$-leptons may provide a sensitive window to physics  beyond 
the Standard Model predictions. The {\tt TauSpinner} algorithm, started with Ref.~\cite{Czyczula:2012ny}, provides a powerful tool to investigate 
characteristics of final states with $\tau$-leptons due to modifications in underlying physics models. 
This is obtained with the help of weights attributed to each event 
from collision data or Monte Carlo generated,
and thus without repeating the detector response simulation with each variant of the physics model.  
An approach where physics assumption 
variation can be introduced with weights is useful for many modern data analysis techniques.
The first version of {\tt TauSpinner} reweighting 
 found its application in the domain of Standard Model measurements \cite{Aad:2012rxl} but 
also for New Physics limits established for simple $2 \to 2$ parton level processes \cite{Banerjee:2012ez}. Later, in \cite{Jozefowicz:2016kvz,Barberio:2017ngd}, 
 {\tt TauSpinner} was found useful for discussion of CP sensitive massively multi-dimensional observables in the frame of Machine Learning 
techniques  \cite{lecun2015deep}.
In a recent paper \cite{Kalinowski:2016qcd}
an extended version 
of {\tt TauSpinner 2.0.0} was presented which now 
includes hard processes featuring tree-level parton matrix elements for production 
of a $\tau$-lepton pair and two jets. 
It was prepared as a tool to be helpful for studying spin effects in processes of Standard Model
and searches of New Physics, like in Refs.~\cite{Aad:2014vgg,CMS:2015mca}.  It found also tempting
applications in the domain of implementation and discussion of
variants for  Standard Model electroweak calculation schemes used in  
simulation programs
 \cite{Kalinowski:2016qcd},  and in 
experimental applications for Standard Model measurements 
\cite{Aad:2012cia,CMS:2016zxv,Cherepanov:2016yvq,Aad:2015vsa}. 

Before discussing {\tt TauSpinner} as a tool for studying observables of 
New Physics let us briefly recall some  {  virtues} of  the  {\tt TauSpinner} algorithm. 
Since $\tau$-leptons cannot be observed directly due to their short life-time with  more than 20 different decay channels, 
each with somewhat distinct signature,  recalculating and reanalyzing  observables involving $\tau$ decays  is time consuming.   
However the $\tau$-lepton spin polarization can be inferred from their decays, contrary to
the case of electron or muon signatures. Spin effects can provide a better insight
into the nature of the underlying physics. Therefore  efforts to explore these phenomena are worth pursuing.  
{\tt TauSpinner} allows  to greatly simplify the task of exploring the experiments' sensitivity.
Evaluation of measurements significance 
due to different New Physics models can be performed with the help of event weights. 
Technical aspects of the algorithm in case of configurations 
with two jets accompanying a $\tau$-lepton pair
have been  covered in \cite{Kalinowski:2016qcd}. 

The purpose of the present paper is to document how the user can apply the {\tt TauSpinner} 
algorithm to his/her physics model. 
To this end, we take as a case study a non-standard  spin-2 object coupled to SM  particles. 
We analyze its  production in proton-proton 
collisions and decay to a $\tau$-lepton pair, {  addressing also the question to what extent  $\tau$ polarization can be exploited to investigate its nature}. We demonstrate how  {\tt TauSpinner} 
can facilitate such studies with the help 
of matrix elements for that model (or any other, provided by the 
user). Corresponding weight  can be calculated and applied to each event of samples with 
full experiment simulation chains. {  As it is practically impossible 
to repeat simulations with detector
response effects included for each  new physics hypothesis, our procedure is beneficial and  may be the only available option for the general use 
despite its limitations.}
{\tt TauSpinner} algorithm can also  be applied 
on measured data events, e.g. in the context of embedded $\tau$ lepton techniques \cite{Aad:2015kxa}.

 The paper is organized as follows: 
Section~\ref{sec:MEtests} and Appendix~\ref{app:HowToUse} provide details of {\tt TauSpinner} 
which were not discussed  in Ref.~\cite{Kalinowski:2016qcd}, or treated very briefly,  
 but are of importance 
 for the case of  New Physics models. 
Section~\ref{sec:ME}  documents details of the tree-level matrix elements used  for the 
calculation of spin-2 object exchange amplitudes which are later passed for weights calculation
in $pp \to \tau \tau\ jj$ events. 
The implemented functionality is based on automatically 
produced {\tt FORTRAN} code from {\tt MadGraph5} package \cite{Alwall:2014hca} similarly as for processes 
of the Drell-Yan--type and of the Standard Model
Higgs boson production in vector-boson-fusion processes (VBF).  We explain details of the modification which we have introduced to the code of 
  {   amplitudes generated  with {\tt MadGraph5} (version  MG5\_aMC\_v2.4.3)}. 
Section~\ref{sec:numerical} and Appendix \ref{app:troubles} are
 devoted to numerical results. First, tests 
for fixed kinematic configurations are recalled.  
Later, definitions of observables are given, and some distributions are 
presented. {   In Section~\ref{sec:spin} numerical results sensitive to $\tau$ polarization are presented taking the single-prong decay  $\tau^\pm \to \pi^\pm \nu$ channel as a spin analyser.
Section~\ref{sec:summary}} summarizes and concludes the paper.

\section{The {\tt TauSpinner} weight $wt_{prod}^{A \rightarrow B}$} \label{sec:MEtests}

{\tt TauSpinner} does not provide methods to generate $pp$ collision events.
Therefore, the necessary input for {\tt TauSpinner} consists of a series of 
events, which could be of a process different than the required one but 
with the same outgoing final states.
The events must contain information on the four momenta of (two) outgoing jets 
and $\tau$-leptons with their decay products, which is necessary for 
the calculation of the hard process matrix elements. 
Flavours of incoming/outgoing partons are determined by the algorithm - there 
is no need to read them from the generated events. The sum over all possible
configurations, weighted with PDFs, is performed. 
On the other hand, the information on the decay products of $\tau$-leptons 
is needed for the evaluation of spin effects.
Using this input, the value of the corresponding matrix elements can be 
calculated on the event by event basis and in particular 
the corresponding spin weight.
As discussed in detail in Ref.~\cite{Kalinowski:2016qcd}, for each event 
\begin{equation}
j_i(p_1) j_j(p_2)\to j_k(p_3) j_l(p_4) \tau^+ \tau^-, 
\end{equation}
where $j$ stands for a quark, antiquark or a gluon, the algorithm calculates the weight
\begin{align}
wt_{prod}^{A \rightarrow B} = & \frac{ \sum_{ijkl}  \frac{1}{\Phi^{i.j}_{flux}} f_i^B(x_1)f_j^B(x_2) 
|M^B_{ijkl}(\{p\})|^2 d\Omega(\{p\})}
     { \sum_{ijkl} \frac{1}{\Phi^{i,j} _{flux}} f_i^A(x_1)f_j^A(x_2)   
|M^A_{ijkl}(\{p\}))|^2 d\Omega(\{p\})} \label{eq:wt_me}
\end{align} 
which represents, for a given phase-space point $(\{p\})=(p_1,p_2,p_3,p_4,p_{\tau^+},p_{\tau^-})$, the ratio due to the matrix element used in the generation of the sample for process (A) and the  matrix elements corresponding to a New Physics model\footnote{It can be also a variant of the Standard Model initialization,
 e.g. distinct electroweak schemes.}  (B).
The evaluation of the weight in Eq.~(\ref{eq:wt_me}) requires the knowledge of contributions from all possible parton level configurations $(ijkl)$  weighted with parton density functions 
$f^{A/B}_{i/j}(x)$ and flux factors $\Phi^{i,j}_{flux}$.
The sums run over both gluons and quark flavours alike.  {  Note however, that the flavour is passed to the 
user provided matrix element routine and flavour dependence can be introduced there.} 
For the details and explanation of a notation used in formula (\ref{eq:wt_me}) we refer to \cite{Kalinowski:2016qcd}. 
For the purpose of calculating  $wt_{prod}^{A \rightarrow B}$ 
we sum over all possible helicity configurations of outgoing $\tau$-leptons.
An event generated for the process (A),  when weighted with $wt_{prod}^{A \rightarrow B}$ becomes an event of  the process (B). 
Spin effects in $\tau$ decays have to be introduced separately, with {\tt TauSpinner} main spin weight $WT$, as explained in Appendix \ref{app:HowToUse}.  

The following details need to be stressed when selecting a suitable process (A) given the target process (B).
For the narrow resonance, like the Higgs state, the value of
matrix elements vary greatly {  with the invariants built from the final state
four-momenta.}
Therefore the numerical stability needs to be kept in mind.
The {\tt TauSpinner} algorithm must reconstruct  invariant mass of 
the resonance with the precision better than 1-2 MeV from the four-momenta of 
final state particles, whose energies may lie in the range of TeV. {  
 Double precision may be needed since otherwise some invariants
may be inappropriately evaluated due to simple computer rounding errors. } 
A user has to assure that 
the reweighting indeed works in the interesting regions of the phase-space. 
In particular, that the phase-space is  populated for both  processes
(A) and (B) with not too massively distinct distributions, and that 
the distribution enhancements due to intermediate resonances or collinear 
or soft singularities have similar (matching) structure. 
Listed above checks require   the hard processes information only.

\section{New Physics model of ($2 \to 4$) process} \label{sec:ME}

{  As a case study we consider a simplified model of a massive gauge singlet spin-2 object $X$ 
coupled to the SM gauge bosons. 
 We use this model to demonstrate  how to prepare and test external 
matrix element 
to be used by {\tt TauSpinner} algorithm.

Scenarios with spin-2 objects  have been} already intensively studied in the literature in the context of  LHC phenomenology \cite{Chiang:2011kq,Artoisenet:2013puc,Frank:2012wh}, though none of the studies  was dedicated to the analysis of $X$ decays into $\tau$ final states. 
Note however that for a  general study of a "Higgs"-like  resonance and its parity in vector-boson-fusion 
processes with a $\tau$ pair as a decay product, experimental results are becoming available 
\cite{Aad:2016nal,Sirunyan:2017khh}.

In Ref.~\cite{Banerjee:2012ez} we studied a Drell-Yan-like production of $\tau$'s through a hypothetical spin-2 object $X$.
Building on our previous work, we study now the $X$ production in the VBF topology, followed by $X \to \tau^+ \tau^-$ decay.
We start by extending the Lagrangian of Ref.~\cite{Banerjee:2012ez} by a set of gauge invariant dimension 5 operators, coupling 
the field $X$ to gauge boson field strength tensors $B$, $W$ and $G$ as
\begin{align}
    \label{eq:lag_gauge_eigenstates}
	\mathcal{L} \ni \frac{1}{F} X_{\mu \nu} \left( \right. & g_{X B B} ~ B^{\mu \rho} B_\rho^{~\nu} + g_{X W W} W^{\mu \rho} W_\rho^{~\nu} \nonumber \\
	& \left. + \, g_{Xg g } G^{\mu \rho} G_\rho^{~ \nu} \right),
\end{align}
where group indices are implicitly summed over (where appropriate).
The parameter $F$, set to 1~TeV, is introduced to keep the coupling constants dimensionless.
Note that we are agnostic on the origin of the state $X$, in particular  we do not claim it is connected to gravity. 
Hence we do not couple it to the entire energy momentum tensor and couplings $g_X$ are kept as free parameters.
This is in contrast to, for example Ref.\cite{Das:2016pbk}, where the $X$ field is coupled to the energy momentum tensor of quantized SM.

\begin{figure}[t!]
	\centering
	\subfloat[]{\includegraphics[width=0.20\textwidth]{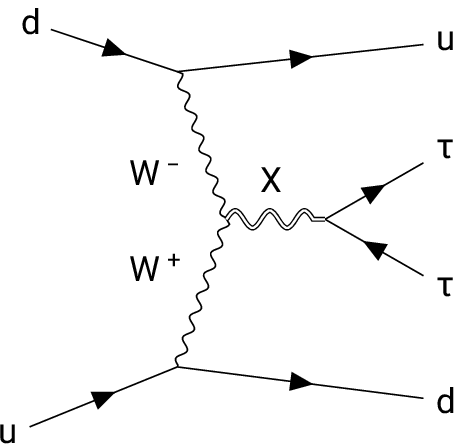} \label{fig:spin2_feynman_diagrams1}}
    \hspace{1cm}
    \subfloat[]{\includegraphics[width=0.20\textwidth]{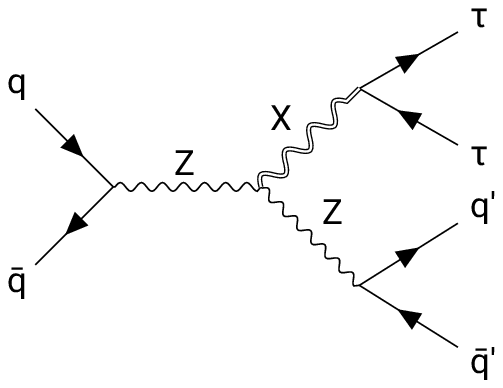} \label{fig:spin2_feynman_diagrams2}}
	\caption{Topologies of Feynman diagrams for $X$ production through its coupling to gauge bosons.
        Similar diagrams, with different combinations of $W^\pm$'s, $Z$'s, photons and quark flavours also exist.
    \label{fig:spin2_feynman_diagrams}
	}
\end{figure}

After the electroweak symmetry breaking,  operators in Eq.~(\ref{eq:lag_gauge_eigenstates}) generate vertices with couplings of $X$ to photons, $W^\pm$'s, $Z$'s and gluons; the explicit formulas for those couplings  can be found in  \cite{Frank:2012wh}.
Since in this work we focus on technical aspects of incorporating  the couplings of $X$ to the EW 
gauge bosons, for numerical tests of the correctness of the Matrix Element implementation, we set 
$g_{XBB} = g_{Xgg} = 0$.
Relevant diagram topologies are shown in Fig.~\ref{fig:spin2_feynman_diagrams}:
for the VBF process (Fig.~\ref{fig:spin2_feynman_diagrams1}) 
and the $X$-strahlung process (Fig.~\ref{fig:spin2_feynman_diagrams2}).

\subsection{Generating matrix-element code using {\tt MadGraph5}}
The extension of the SM by spin-2 field coupled to the gauge fields as in 
Eq.~(\ref{eq:lag_gauge_eigenstates}), including also coupling of the $X$ field  
to quarks and $\tau$-leptons from \cite{Banerjee:2012ez}, is encoded into a \texttt{FeynRules}
~\cite{Alloul:2013bka} model.
The \texttt{FeynRules} model file, together with its \texttt{UFO} output \cite{Degrande:2011ua}, 
is available 
in supplementary materials of the \texttt{arXiv} version of this reference.
The \texttt{UFO} model is used to generate squared matrix elements using \texttt{MadGraph5}, employing the spin-2 support of 
the  \texttt{HELAS} library \cite{Hagiwara:2008jb}. 
This is done with the following set of commands
\begin{itemize}
	\item[(a)] \texttt{import model spin2\_w\_CKM\_UFO}
	\item[(b)] by default, ``multiparticles'' containers already include all massless partons\\
	\texttt{p = g u c d s u\char`\~ ~c\char`\~ ~d\char`\~ ~s\char`\~}\\
	\texttt{j = g u c d s u\char`\~ ~c\char`\~ ~d\char`\~ ~s\char`\~} 
	\item[(c)] generate spin 2 matrix elements \\
	\texttt{generate p p > j j x QED<=99 QCD<=99 NPgg<=99 NPqq<=99 NPVV<=99, x > ta+ ta-}
	\item[(d)] write the output to disk in \texttt{MadGraph}'s standalone mode using\\ \texttt{output standalone "directory name"}
\end{itemize}
\texttt{NPgg}, \texttt{NPqq} and \texttt{NPVV} parameters control the maximum 
number of $g_{Xgg}$,  $g_{Xq\bar{q}}$ and  $g_{XWW}$, $g_{XBB}$ couplings, respectively. 
Limiting them by 99 effectively means that their number is not restricted.
The model includes the CKM matrix in the Wolfenstein parametrization.
As was stated above, for numerical tests we restrict ourselves setting $g_{XBB} = g_{Xgg} = 0$, though we stress again 
that the matrix element, coded as an example user process, 
contains all of them, see Appendix \ref{app:troubles} for actual initialization of coupling constants.
\begin{table*}[htb]
	\begin{center}
  \caption{
 	List of implemented processes contributing to the spin-2 X particle production, grouped into categories which differ by flavours of incoming partons. 
 	For each category, the names of {\tt FORTRAN} files calculating squared 
 	matrix elements, for given flavour configuration of incoming partons, 
 	are given in the second column. 
 	Examples of processes in each category are given in the last column.
 }\label{tab:processes}
  \begin{tabular}{lll}
  \hline\noalign{\smallskip} 
 Category of               &  Corresponding {\tt FORTRAN} files    & Processes       \\ 
 Matrix Elements           &                                 &                 \\ 
  \noalign{\smallskip}\hline\noalign{\smallskip}
   { (1)}                  & $GG\_S2.f$                         & $g g \to \sum_{f} q_f \bar q_f X$ \\
\noalign{\smallskip}\hline\noalign{\smallskip}
   {  (2) }                 & $GC\_S2.f$, $GU\_S2.f$     & $g q_f  \to g q_f X$\\
\noalign{\smallskip}\hline\noalign{\smallskip}
 { (3) }                    & $GCX\_S2.f$, $GUX\_S2.f$     & $g \bar q_f  \to g \bar q_f X$\\
 \noalign{\smallskip}\hline\noalign{\smallskip}
   {  (4) }                 & $DD\_S2.f$,   $UD\_S2.f$,  $UU\_S2.f$,     & $q_{f_1} \ q_{f_2}\ (\bar q_{f_1} \ \bar q_{f_2})   \to q_{f_1} \ q_{f_2} (\bar 
                           q_{f_1} \ \bar q_{f_2}) X$ \\
                           & $CC\_S2.f$, $CS\_S2.f$,            &   \\ 
                           & $SS\_S2.f$  $CD\_S2.f$,             &  \\
                           & $CU\_S2.f$, $SD\_S2.f$, $SU\_S2.f$             &   \\
\noalign{\smallskip}\hline\noalign{\smallskip}
   {  (5) }                &  $DDX\_S2.f$, $UDX\_S2.f$,  $UUX\_S2.f$      & $ q_{f_1} \ \bar q_{f_2}\ (\bar q_{f_1} \ \bar q_{f_2})   \to q_{f_1} \ \bar q_{f_2} (\bar q_{f_1} \ \bar q_{f_2})X $  \\
                           & $CCX\_S2.f$, $CSX\_S2.f$, $DCX\_S2.f$,             &    \\ 
                           & $SCX\_S2.f$, $SSX\_S2.f$,  $UCX\_S2.f$,           &   
                           $q_{f_1} \ \bar q_{f_2}\ (\bar q_{f_1} \ \bar q_{f_2}) \to ggX $    \\
                           & $CDX\_S2.f$, $CUX\_S2.f$, $SDX\_S2.f$,           &  \\
                           & $SUX\_S2.f$  & \\

\hline
\end{tabular}
\end{center}
\end{table*}
\subsection{Integrating matrix-element code into {\tt TauSpinner} example }\label{sec:MadGraph}

The matrix element code is based on  automatically produced {\tt FORTRAN} 
subroutines by {\tt MadGraph5} package, similarly as it has been done for processes 
of the Drell-Yan--type and of the Standard Model
Higgs boson production in vector-boson-fusion (VBF)/Higgs-strahlung  processes \cite{Kalinowski:2016qcd}.  
In the spin-2 case they have been also manually modified and adapted to  avoid name clashes. 
This technical complication is a consequence of the fact that  {\tt C++} user 
function for the spin-2 matrix element calls 
{\tt FORTRAN}  code created by {\tt MadGraph5}. 
 We therefore can not profit from {\tt namespace} functionality of {\tt C++} as a  natural solution to this problem. 
Some name changes are necessary, as explained below. The corresponding code is stored in the
directory {\tt TauSpinner/examples/example-VBF/SPIN2/ME}.

The generated codes for the individual sub-processes are 
grouped together into subroutines, 
depending
on the  flavour of initial state partons, and named accordingly. For example, 
\begin{verbatim}        
SUBROUTINE  DCX_S2(P,I3,I4,H1,H2,ANS)
\end{verbatim}
encompasses the $X$ production processes initiated by the $d\bar c$ partons.  We follow 
our previous convention \cite{Kalinowski:2016qcd} where symbol {\tt X} in the subroutine or internal function name after the letter {\tt U,D,S} or {\tt C} means that the corresponding parton is an antiquark, 
i.e. {\tt UXCX} corresponds to processes  initiated by $\bar u\bar c$ partons, while {\tt GUX} to 
processes initiated\footnote{{\tt X} in this context should not be confused with the spin-2 field $X$.}  by $g\bar u$.
The {\tt S2} stands explicitly for the production of  spin-2 $X$ state.  
The input variables are: real matrix {\tt P(0:3,6)} for four-momenta of incoming and outgoing
particles, integers {\tt I3,I4} for the Particle Data Group (PDG) identifiers for final state parton flavours and 
integers {\tt H1,H2}  for  the
outgoing $\tau$ helicity states.
Before integrating these subroutines into the {\tt TauSpinner} program, a number of modifications have 
been done for the following reasons:
\begin{itemize}
\item[a)] Since {\tt MadGraph5} by default sums and averages over spins of incoming and outgoing particles,
 while we are interested in $\tau$ spin states,  
the generated codes have to be modified to keep track of the $\tau$ polarization, 
i.e. indices/helicities {\tt H1} and {\tt H2}.
\item[b)] Moreover, since  the subroutines and internal functions  generated by {\tt MadGraph5} have the same 
names for all sub-processes, namely {\tt SMATRIX(P,ANS)}, the names had to be changed to be 
unique.
As an example, the subroutine name for the subprocess 
$u\bar d\to c\bar d \, X,\, X\to\tau^+\tau^-$  was changed to 
{\tt  UDX\_CDX\_S2(P,H1,H2,ANS)}. These subroutines will be called by 
subroutine \\ {\tt UDX\_S2(P,4,-1,H1,H2,ANS)}.
\item[c)] For  a pair of final-state parton flavours $k \ne l$, the 
{\tt MadGraph5} generated codes have been obtained 
for a definite ordering
$(k,l)$, but not for  $(l,k)$, to reduce the number of generated configurations.
When  {\tt TauSpinner} is invoked, 
the  configuration of outgoing partons is unknown and it takes into account 
both possibilities: thus a compensating factor  $\frac{1+\delta_{l,k}}{2} $ has to be introduced due to  the way of organizing the sum 
in Eq.~(\ref{eq:wt_me}) and in  Ref.~\cite{Kalinowski:2016qcd}.
\item[d)] For calculation of matrix elements {\tt MadGraph5} is using ALOHA 
functions \cite{deAquino:2011ub} stored in 
{\tt FORTRAN} subroutines. Since some of these functions 
have originally names identical to functions in the  {\tt TauSpinner} 
source code for the implementation of the Standard Model VBF\slash{}Higgs-strahlung  production,   
names of those functions have to be modified also  to avoid any name conflicts. 
Therefore ALOHA functions  stored in 
 \texttt{TauSpinner\slash{}examples\slash{}example-VBF\slash{}SPIN2\slash{}ME/Spin2\_functions.f}
are changed by  
adding "{\tt \_S}" suffix to the original names of subroutines,  for example \texttt{FFV4\_0} is 
changed to \textnhtt{FFV4\_0\_S}.
\end{itemize}

Table \ref{tab:processes} summarizes the naming convention for the files.
At the parton level each of the incoming or outgoing partons can be one of 
the flavours: $\bar c\ \bar s\ \bar u\ \bar d\  g\ d\ u\ s\ c$, 
with Particle Data Group (PDG) identifiers: 
  -4, -3, -2, -1, 21, 1, 2, 3, 4 respectively.
For processes with two incoming partons,  two outgoing $\tau$-leptons 
and two outgoing partons, there are
$9^4$ possibilities, most of them evaluating to 0 {   or obtainable} one from 
another, by relations following from  CP symmetries 
and/or  permutations of incoming and/or outgoing partons.

For each point in the parton level phase-space, consisting of all incoming and outgoing four momenta as well as their flavours, 
depending on the user choice, one of two variants 
of processes (i.e. pairs of matrix elements)  may 
be used by {\tt TauSpinner} executable.
That is: the Drell-Yan variant (standard, and user provided  New Physics 
matrix elements) or Higgs-like variant  (again standard, and user provided one%
\footnote{The prototype is implemented in the example, see Appendix \ref{sect:MainProgram}.}%
).

Certain limitations need to be kept in mind. In practice, 
it  is simply impossible to obtain statistically significant distribution 
of weighted events for the 
particular model under study in the region of phase-space where original sample is sparse or possibly no events are present at all.
In particular, the mass and width of the  Higgs-like resonance need to coincide
 (be close) to those of the Higgs.
Also, the algorithm is expected to be used in regions of the phase-space where kinematic
distributions of the original and New Physics models are not massively different. 


\section{Tests of implementation of external matrix elements} 
\label{sec:numerical}
Once the user-provided external matrix elements 
are prepared, numerical 
tests are necessary if it indeed  has been implemented properly into the  {\tt TauSpinner} 
environment.
In the following we discuss such  tests, using
spin-2 matrix elements of $Xjj$ production as an example. We start from 
the technical one and continue with more physics oriented ones.
Finally we will demonstrate limitations of the method. 

\begin{figure}[h!]
 \begin{center}                              
{
 \includegraphics[width=8cm,angle=0]{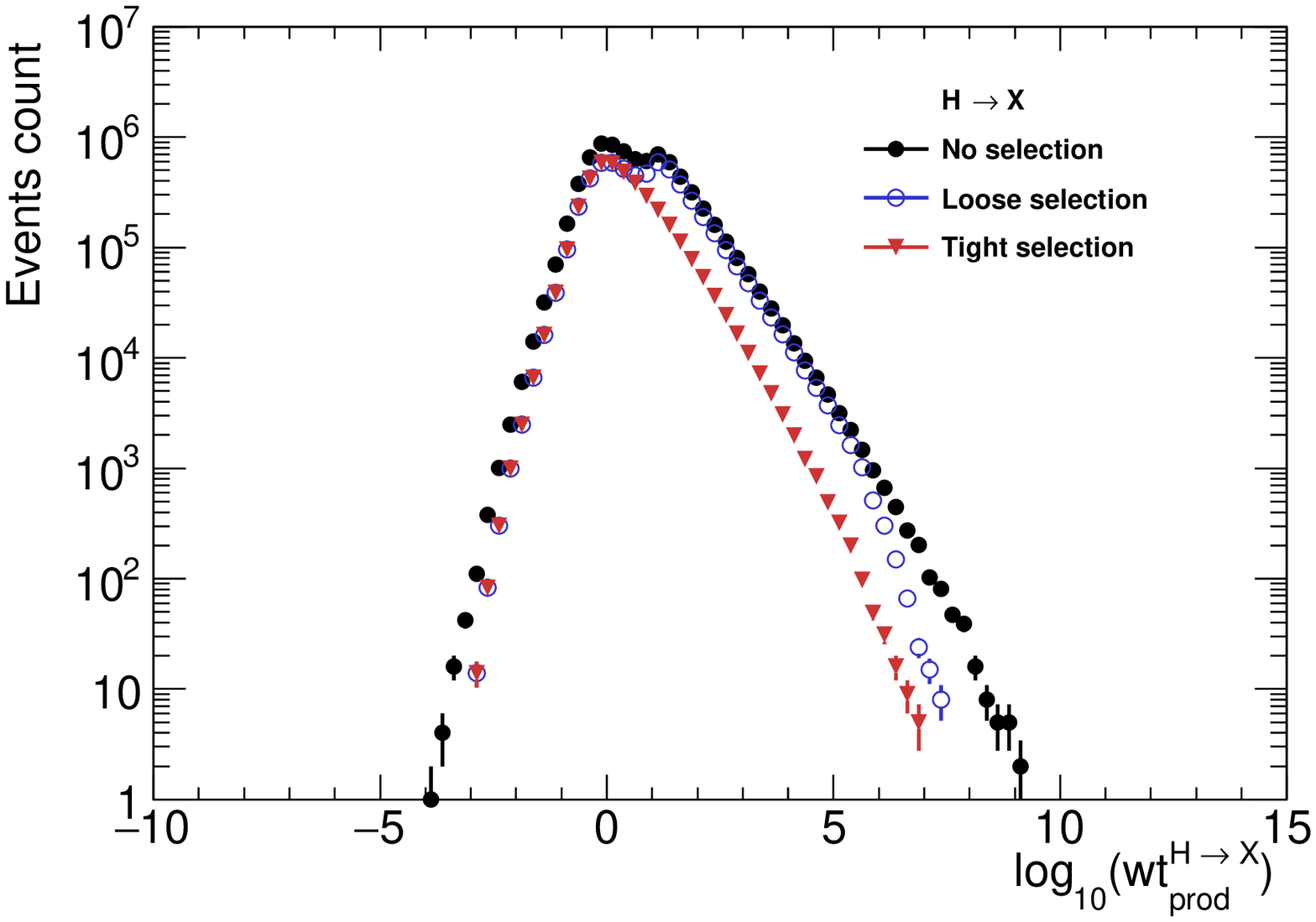}
 \includegraphics[width=8cm,angle=0]{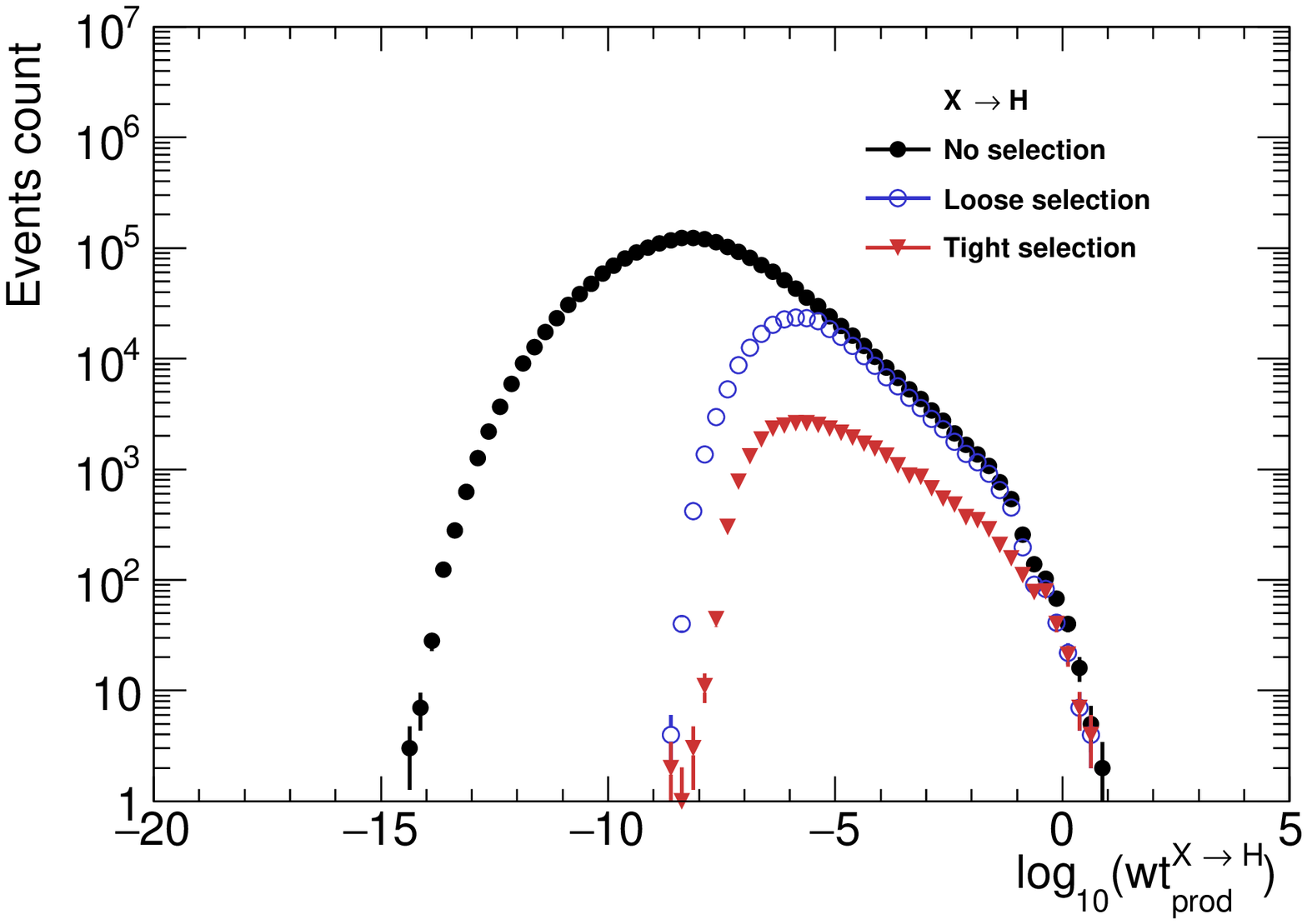}
 }
\end{center}
\caption{Weight distribution for $H$ sample reweighted to $X$ (top panel) and for the $X$  sample reweighted to $H$ (bottom panel). 
If the distribution featured a long tail extending to high weights, it would indicate a problem with reweighting in regions
 of the 
phase-space where the ratio of  the matrix element (B) with respect of the one of  the original sample (A) is too large  
in comparison  to the typical event.   }\label{fig:one}
\end{figure}

\subsection{Test of matrix elements using fixed kinematic configuration}
For checking the consistency of the implemented codes generated with {\tt MadGraph5} and modified as explained in section \ref{sec:MadGraph},
 we have chosen a single event with fixed kinematic configuration at the parton level. 
We have calculated the matrix elements squared for that event and for all 
possible helicity and parton flavour configurations, using the code implemented as user example. We compared results with the numerical values obtained 
directly from {  {\tt MadGraph5}}. The agreement of at least 6 significant digits has been confirmed. 

\subsection{Tests of matrix elements using series of generated events}\label{sec:kinematical}
As  further tests of the internal consistency of external matrix element 
implementation, we have explored the reweighting procedure 
by comparing a number of kinematic distributions obtained directly or reweighted with $wt_{prod}^{A \rightarrow B}$ from  
series of 10M  events generated by {\tt MadGraph5} 
for $X$ particle 
 and  Higgs boson. 
Samples were generated for {\tt pp} collisions at 13 TeV using  
{\tt CTEQ6L1} 
 PDFs. The mass of both $X$ particle and Higgs boson  was set to 125~GeV and the 
width to 5.75 MeV.
The details of cuts and {\tt MadGraph5} initialization
 used for the sample generation are given  in Ref.~\cite{WebPageSoinner2j}.  
On the generated events the following further selections
were applied: $m_{jj\tau\tau}$ $<$ 1500 GeV,  $p_{T}^{\tau\tau}$ $<$ 600~GeV
and $ m_{jj} <800$ GeV (loose selection) 
{  to eliminate excessive weight regions of the phase space}, 
or eliminating also $Z\to jj$ or $W\to jj$ 
resonance peaks $100 < m_{jj}$ $<800$ GeV ({  tight} selection).

 Before commenting on the actual results let us point to  the size of statistical errors\footnote{ 
The statistical errors of all histograms, including the ones using weighted events,   were evaluated  by the standard algorithms of the root library \cite{Antcheva:2009zz}.}  
which reflect comparability of the $H$ ( process A) and $X$ (process B) samples.
Errors  are always larger than what could be expected from weight-one samples of the similar size.
This effect can be understood better with the  weight distributions shown in Fig.~\ref{fig:one}.
In both cases of reweighting: from $H$ to $X$ (top panel) and $X$  to $H$ (bottom panel), one can observe a 
 constant slope on this double logarithmic plot with clear sharp upper end.
With such spectrum of weights statistically sensible 
calculation of the cross sections
and distributions may be  still possible. If a tail of events with \\

\onecolumn
\begin{figure}
 \begin{center}                              
{
   \includegraphics[width=7.5cm,angle=0]{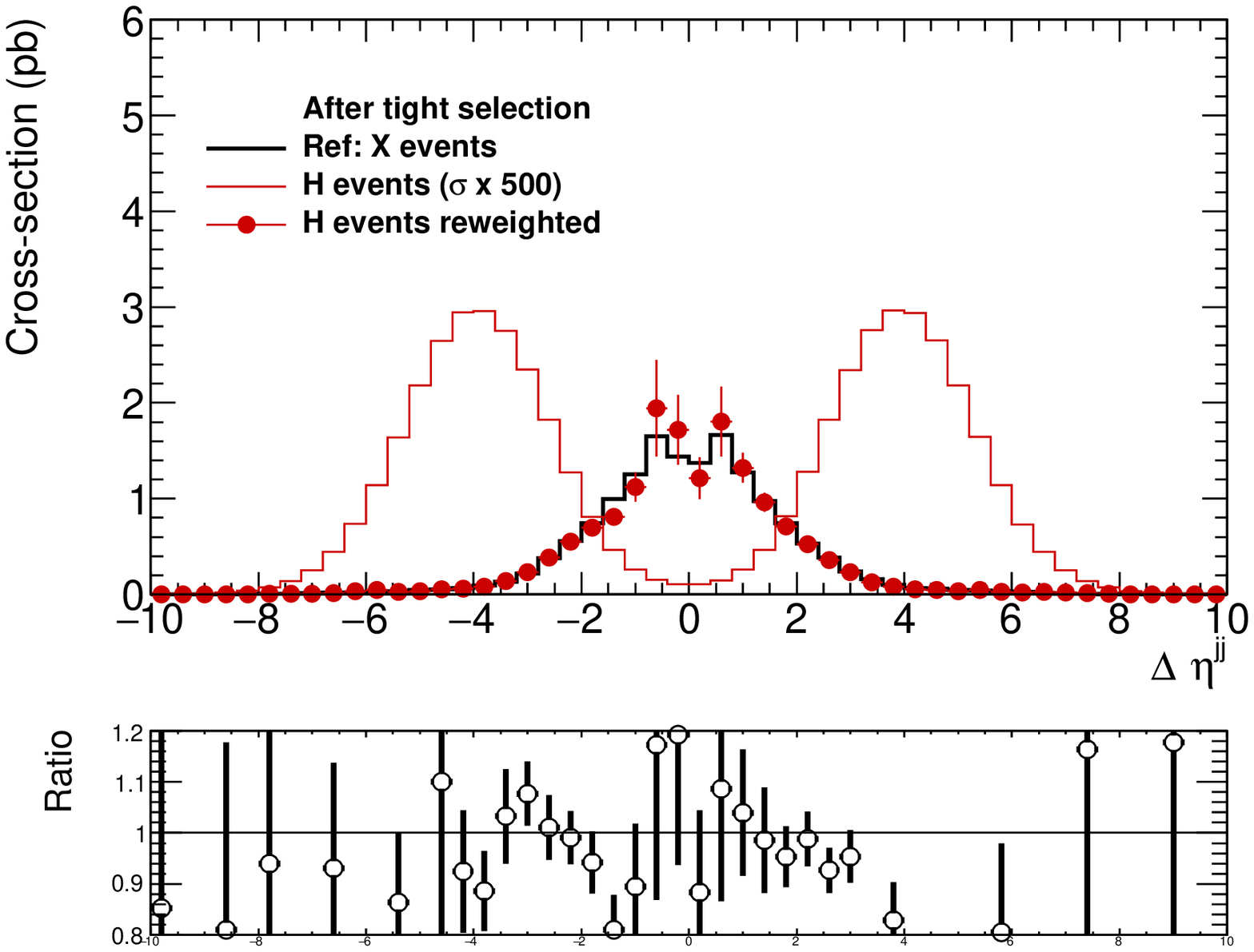}
   \includegraphics[width=7.5cm,angle=0]{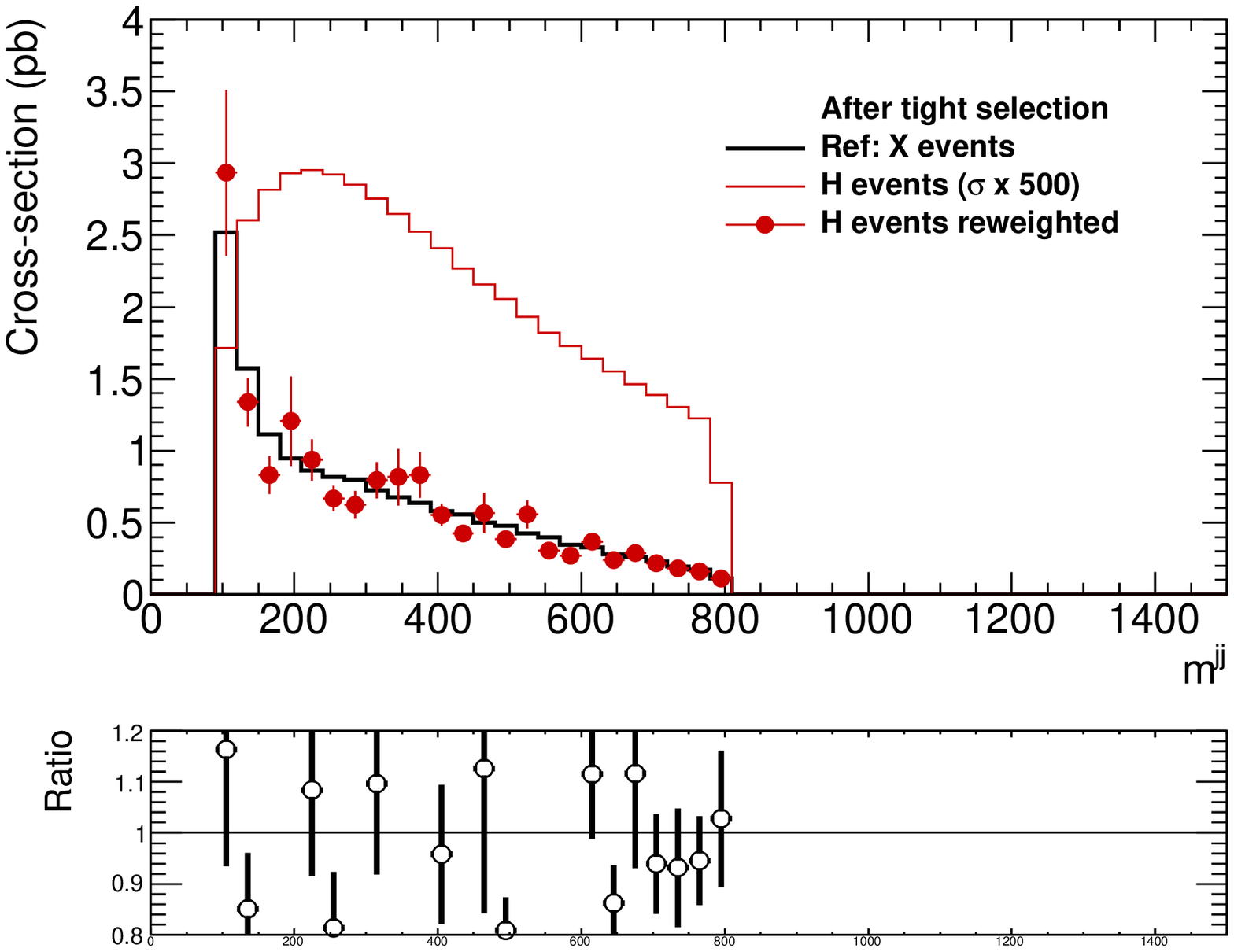}
   \includegraphics[width=7.5cm,angle=0]{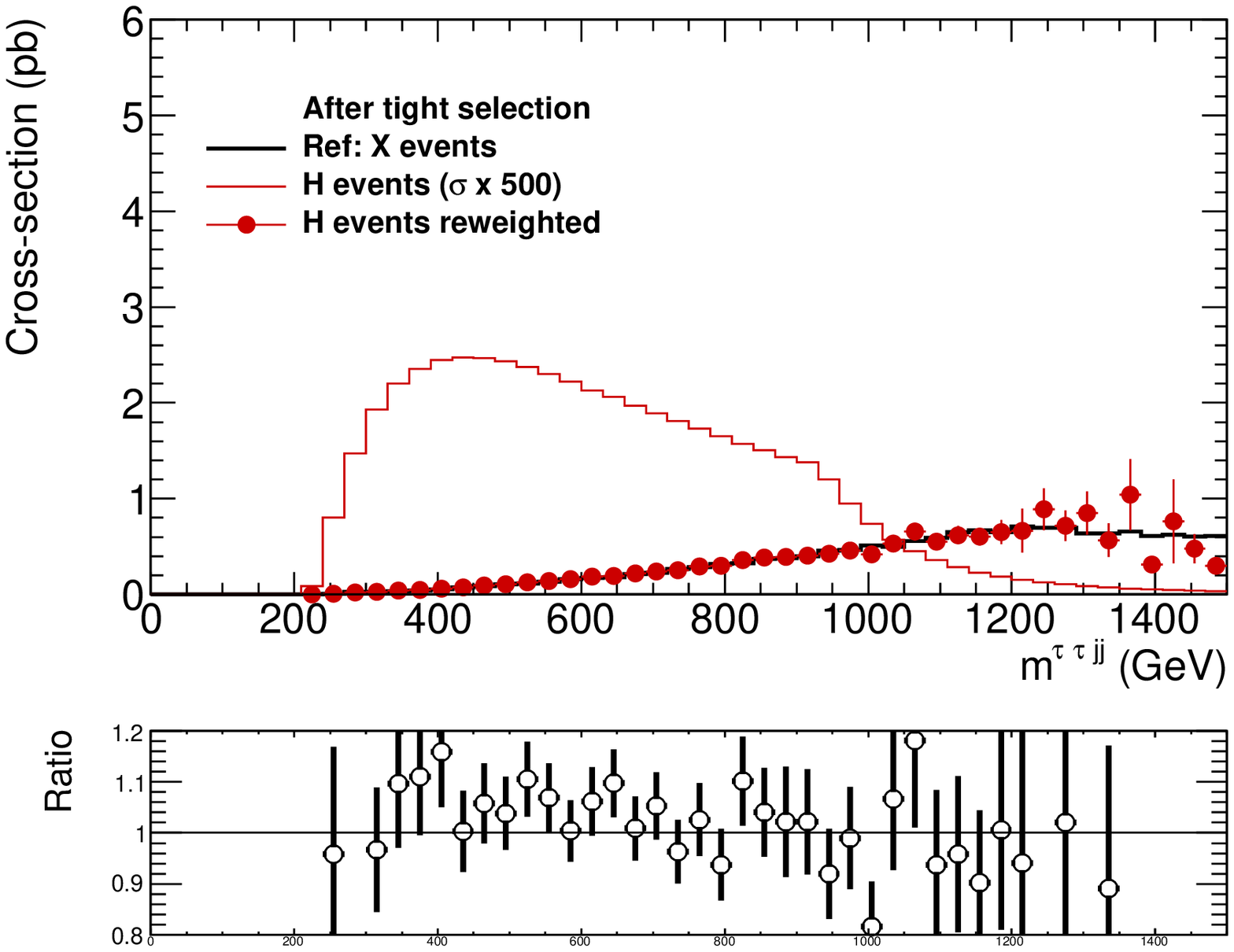}
   \includegraphics[width=7.5cm,angle=0]{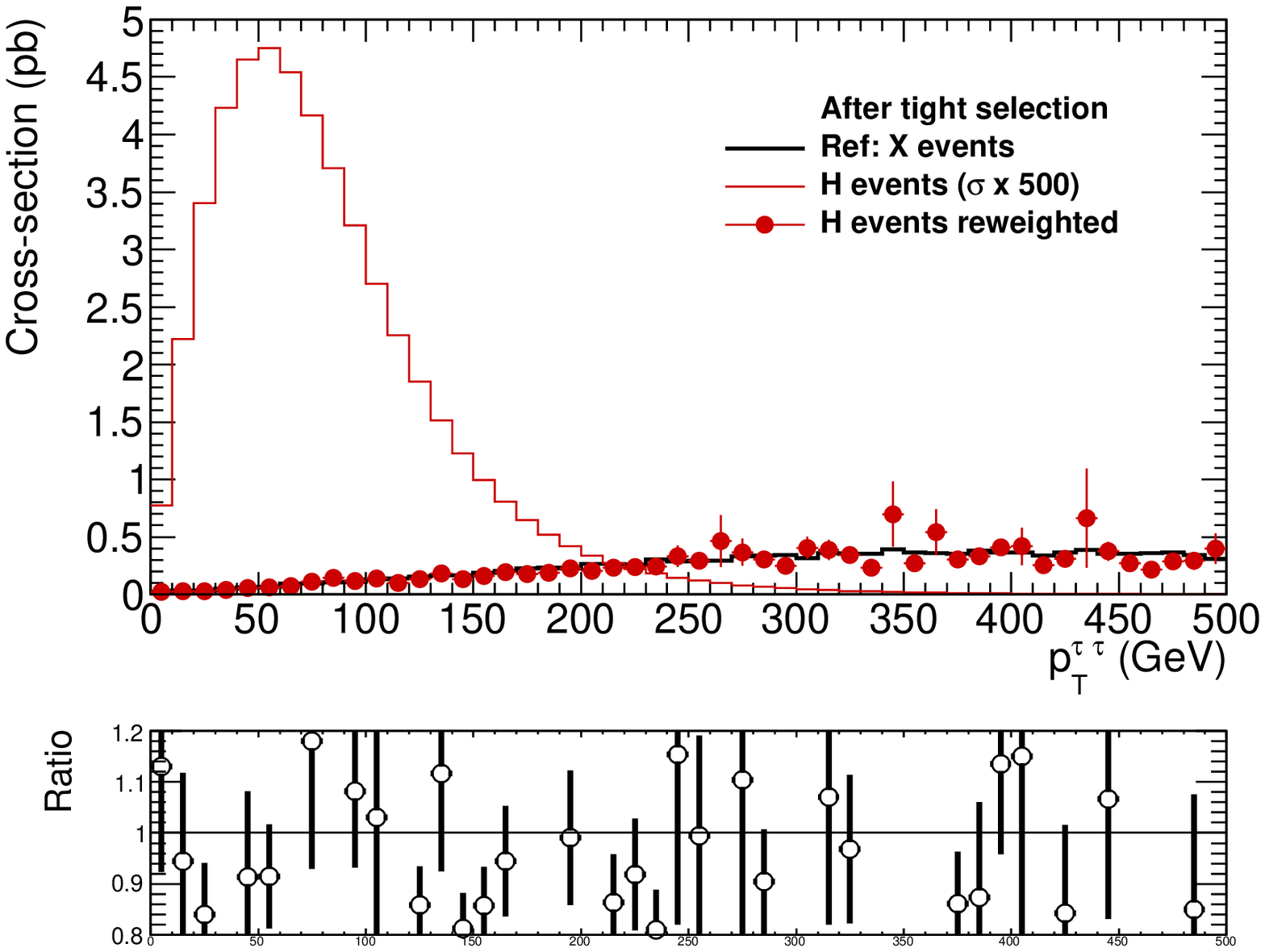}
}
\end{center}
\caption{The $H$ sample reweighted to the $X$ and compared with the  $X$ sample.
 The $H$  and $X$ widths  are of 5.75 MeV. 
Selection cuts: Invariant mass of outgoing particles $m^{\tau\tau jj} <$ 1500~GeV, 
invariant mass of jets system $100< m^{ jj}<$ 800 GeV and $p_{T}^{\tau\tau}$ $<$ 600 GeV.
Variables on the x-axes as explained in  Section \ref{sec:kinematical}. \label{fig:four} }
\end{figure}
\begin{figure}
\begin{center}                              
{
   \includegraphics[width=7.5cm,angle=0]{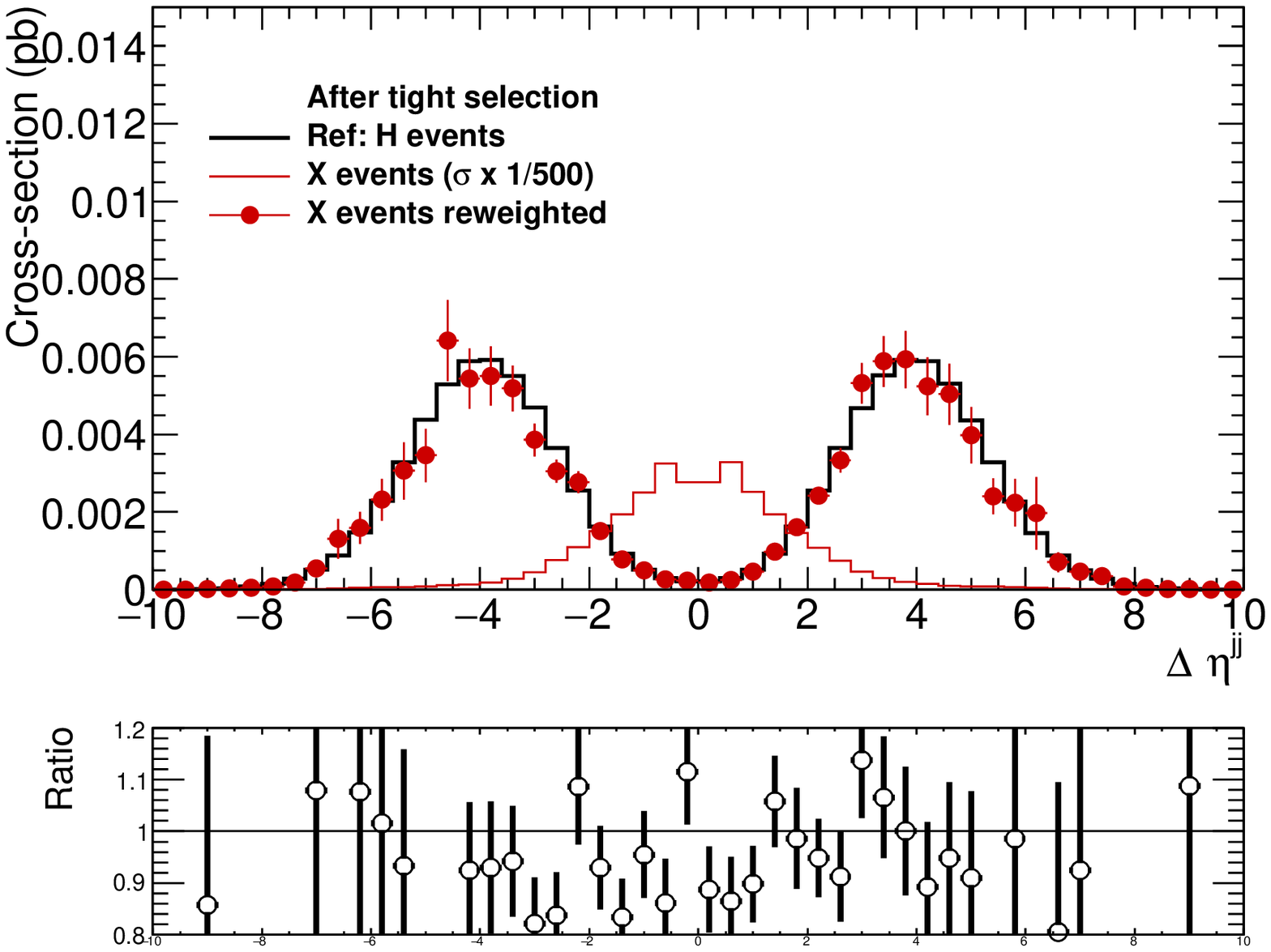}
   \includegraphics[width=7.5cm,angle=0]{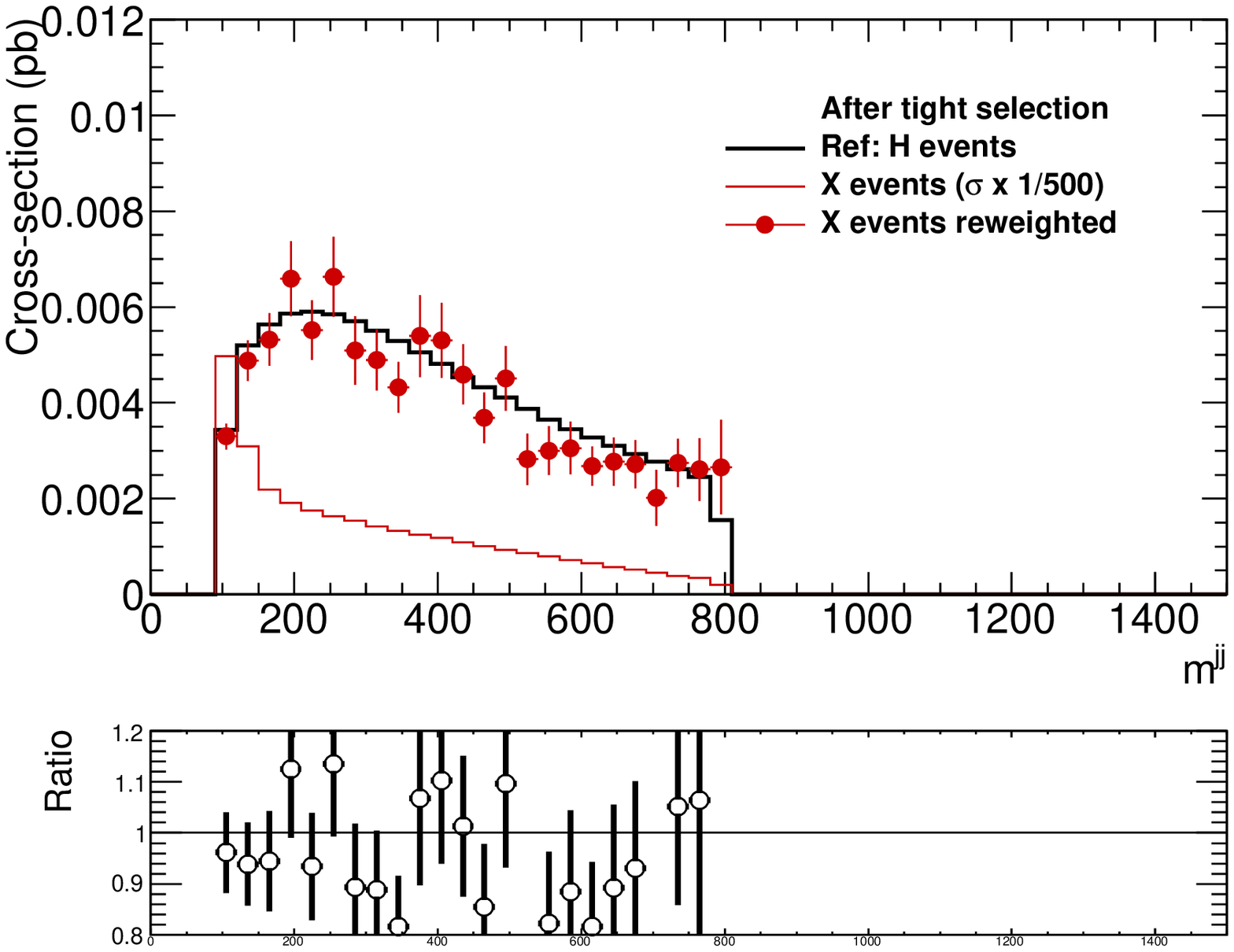}
   \includegraphics[width=7.5cm,angle=0]{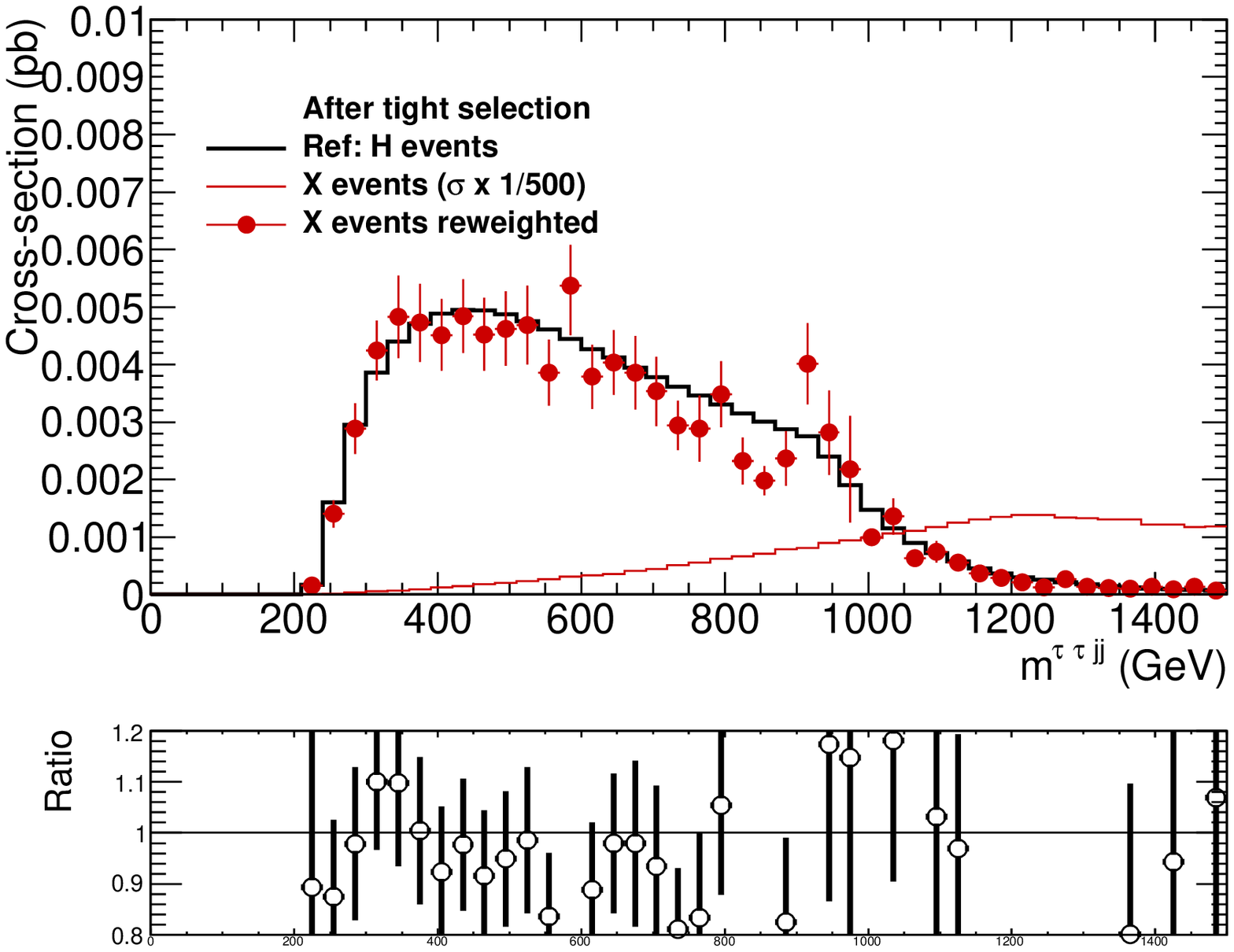}
   \includegraphics[width=7.5cm,angle=0]{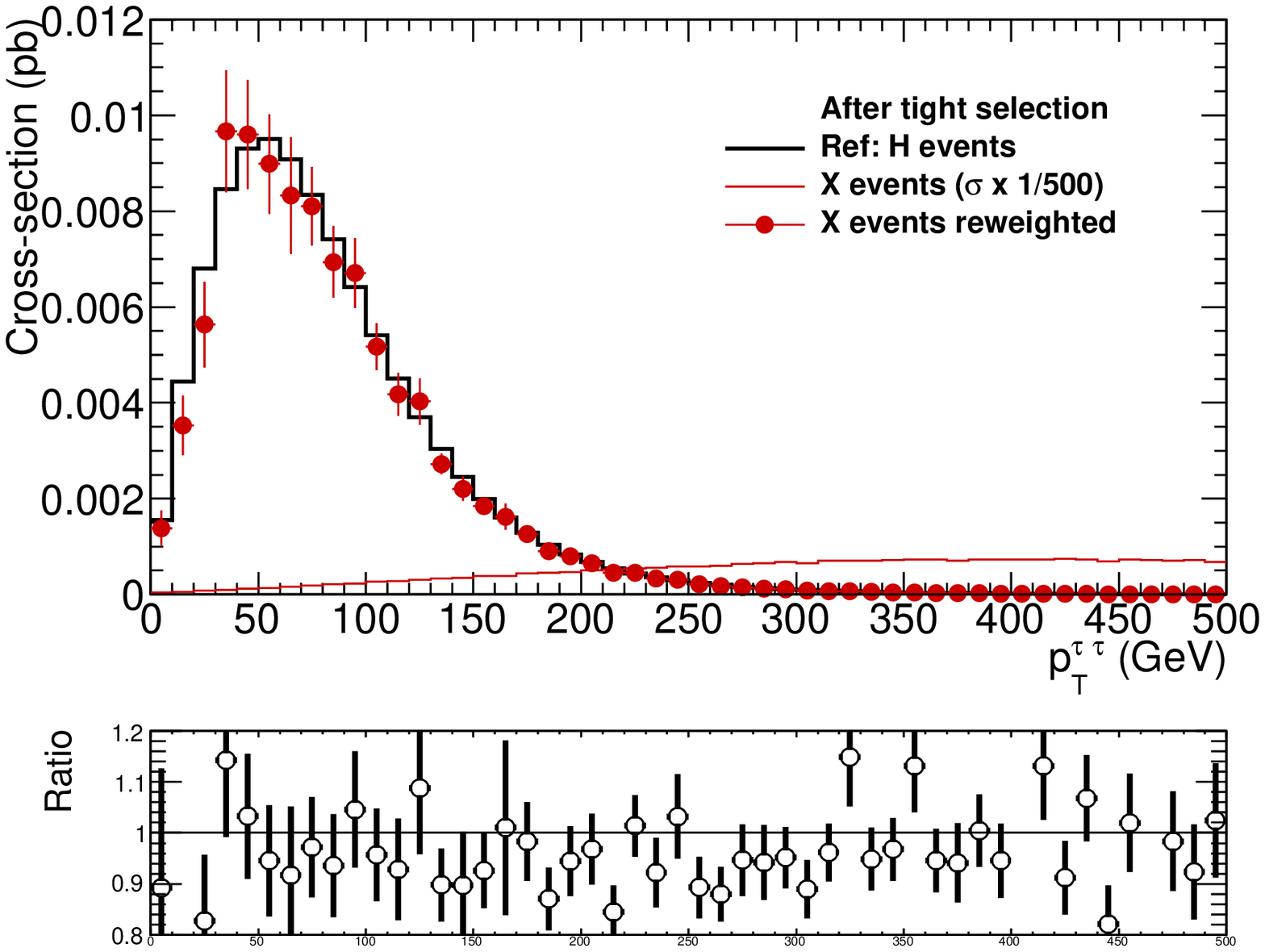}
}
\caption{The $X$ sample reweighted to the $H$ and compared with the $H$ sample. 
	The $H$  and $X$ widths are of 
	5.75 MeV. Selection cuts: 
	invariant mass of outgoing particles $m^{\tau\tau jj} <$ 1500~GeV, 
	invariant mass of jets system $100 <m^{ jj}<$ 800 GeV and  $p_{T}^{\tau\tau}$ $<$ 600 GeV.
	Note that statistical errors for the distributions obtained with 
	reweighting (red points) are much larger than for the case of Fig.~\ref{fig:four}. This is predominantly 
	due to small acceptance of $X$ sample: 1.7\% only. 
	But the agreement with the reference distribution (black histogram) remains within statistical 
	fluctuation (dominated by large weight events). Variables on the x-axes
	as explained in Section \ref{sec:kinematical}.\label{fig:five}}
\end{center}
\end{figure}

\begin{table*}[t]
\begin{center}
 \caption{Cross sections for the generated $H$ production process and after its reweighting to the $X$ 
	production ($H\to\tau\tau$ block), and for the generated $X$ production and after its reweighting to $H$ 
	production ($X\to\tau\tau$ block); acceptances with no, loose or tight selections applied  for generated 
	and reweighted event samples are also shown.}\label{tab:xsect}
  \begin{tabular}{llccc}
  \noalign{\smallskip}\hline\noalign{\smallskip}
 Events     &             &  No selection   & Loose selection  & Tight selection       \\ 
  \noalign{\smallskip}\hline\noalign{\smallskip}
             & Acceptance  & 100\%       & 73.8\%      & 49.0\%   \\
  $ H \rightarrow \tau \tau$  & $\sigma$ [pb] (H)  &[2.033$\pm$0.064] $10^{-1}$  & [1.501$\pm$0.062] $10^{-1}$ & [1.004$\pm$0.045] $10^{-1}$\\
             & $\sigma$ [pb]  ($H  \rightarrow X$) &[9.097$\pm$1.270] $10^{+2}$  & [1.187$\pm$0.038] $10^{+2}$ & [1.517$\pm$0.066] $10^{+1}$  \\
   \noalign{\smallskip}\hline\noalign{\smallskip}
             & Acceptance  & 100\%       & 13.0\%      & 1.71\%   \\
  $ X \rightarrow \tau \tau$  & $\sigma$ [pb] (X)  &[9.097$\pm$0.0029] $10^{+2}$ & [1.178$\pm$0.001] $10^{+2}$ & [1.544$\pm$0.004] $10^{+1}$ \\
             & $\sigma$ [pb]  ($X  \rightarrow H$) &[2.023$\pm$0.0474] $10^{-1}$ & [1.478$\pm$0.031] $10^{-1}$ & [ 9.75$\pm$0.309] $10^{-2}$  \\
   \hline
\end{tabular}
\end{center}
\end{table*}

\twocolumn

\noindent ever higher weights would continue
to form when increasing size of samples,  statistical errors would never decrease.
{  This happens, for example, if in some sub-dimensional-manifold of the phase space the 
matrix element has a zero. Then with increasing statistics, events closer and closer to this zero are
 generated, and feature larger and larger weights. Even though contribution of 
such events
to the weighted distribution is formally finite and integrable, the error estimate of the 
Monte Carlo generated  distribution 
will not get reduced with the increasing statistical sample. }

The tests were performed on a set of kinematic distributions:
 pseudorapidity of outgoing parton $j$,
rapidity of $\tau\tau$ and $jj$ systems,
 invariant mass of $\tau\tau$ system,
pseudorapidity of $\tau\tau$ system,
opening angle between jets, 
opening angle between $\tau$'s,
angle between incoming parton and outgoing parton in the rest frame of jets and
angle between resonance and outgoing parton in the rest frame of jets.

Plots for all these variables  can be found on the web page \cite{WebPageSoinner2j}.
Here, in Fig.~\ref{fig:four} and Fig.~\ref{fig:five}, we present only   plots for: the  difference of jet's rapidities $\Delta\eta^{jj}$, 
the invariant mass of the jet pair $m^{jj}$,
the  transverse momentum of $\tau$ pair $p_T^{\tau\tau}$ and finally  the  invariant mass of $\tau$-pair and 
jet-pair  combined $m^{\tau\tau jj}$.
In each plot 
the distribution {\tt Ref}, for the reference process, is shown as 
a black histogram  while the red histogram is the original distribution of generated events 
 which are reweighted using  {\tt TauSpinner}   $wt_{prod}^{A \rightarrow B}$ 
weight to obtain  the 
 distribution represented by the  red points with 
error bars. For the test to be successful, the red points should follow 
the black histogram; the ratio of {\tt Ref} and reweighted distributions 
is shown in the bottom panel of each figure.

In both  Figs. \ref{fig:four} and  \ref{fig:five}, the reweighted distributions follow the {\tt Ref} 
histograms. When reweighting of $X$ to $H$ (see  Fig.~\ref{fig:five}), the distributions 
feature larger statistical errors than in the case of $H$ to  $X$  reweighting (Fig.~\ref{fig:four}).
This is simply because tight selection cuts leave only 1.7\% of $X$
 events due to eliminating configurations with small $m_{jj}$.
For  some bins the reweighted distribution lies below target (black) distribution, 
whereas the ones with big errors tend to lie
above. {  If similar feature  appears when sample size
          is increased it points to the possibility that original distribution had a zero 
          along some hyperspace.  Nevertheless, if in distributions normalized to cross section 
          the neighbouring bins have no deficit of content, then the reweighting 
          algorithm can be still used.}

The tests validating reweighting algorithm are completed with the ones monitoring
 overall normalizations (integrated cross sections).
For our samples and initializations, the resulting cross sections are shown in Table
\ref{tab:xsect}. Reasonable agreement between cross sections obtained from the {\tt MadGraph5} calculation and 
with reweighting was obtained, see Table \ref{tab:xsect} where the first line in the $H\to\tau\tau$ 
block should be compared to the second line in the $X\to\tau\tau$ block and vice-versa.  
Such a study has to be repeated for each new matrix element implemented and 
whenever selection cuts are changed sizably.

\subsection{On reliability of the TauSpinner reweighting approach
\label{sec:reliability}}

The TauSpinner reweighting method is atypical
compared to methods used in other tools,  like REPOLO \cite{Schissler:2014nga}
  PHYTIA  \cite{Mrenna:2016sih},  SHERPA. \cite{Bothmann:2016nao} or 
 MadGraph  \cite{Artoisenet:2010cn}.
 Let us explain what are 
the advantages and disadvantages behind such a choice. 

The advantage of our method is that it does not assume any knowledge of the initial and outgoing partons 
and tau leptons beyond their four momenta. Therefore it can be applied 
directly to the experimental data, e.g. of the  embedded $\tau$ 
samples. 
We have demonstrated that our reweighting method 
is reliable for the hard process matrix elements convoluted with PDF's. 
The disadvantage is that it does 
not address the issue  that both the parton shower and 
hadronization  do depend on color configuration as well as on flavours of 
partons. Once event  is reweighted, the reshuffle between categories of
different color and/or flavour content takes place, inevitably leading to biases.

In experiment simulation production files  \cite{Aad:2010ah} 
color  information  for the so called truth entries is not stored. 
Even  in the data formats prepared  
and agreed on by the community \cite{community}, such 
information, at best,  consists of a connected tree, navigation inside of which retains 
information on the event history including the parents of unstable particles. There is an 
important caveat here: the event generators are modeling quantum processes, and 
the event record has the structure of a classical decay chain. It is inevitable that 
compromises must be made and difficulties can arise from an over-literal interpretation 
of the tree structure. 
For the color it means that at best the so called flow approximation is pre-imposed.
Even for such partial information,  there are no detailed commonly accepted
rules how it should be stored, see for example Section 2.3 in  \cite{Dobbs:2001ck}
or Section 4.4.1 in the HepMC manual \cite{hepmcmanual}
 or ~\cite{Alwall:2006yp}.

In practice, in experiment production files of detector response simulated events,
information on intermediate quantum states is generally  not available.
Usually only the 4-momenta of partons and their flavours are stored.  
We are not in a position to affect these  experiments choices. 
Multitude of arguments  have been raised for such choices, including the fact that distinct generators
prepared by theorists, provide such information in different manner, or that it makes 
data files unnecessary large. 
We can only address the question if any useful solution for reweighting 
may be designed\footnote{
Note that for spin effects we include in our re-weighting not only production matrix element,
but the ones of $\tau$'s decays as well.
} and what kind of restrictions it implies  have to be kept in mind.

There are two simulation steps which depend on hard process configurations
of flavours and colors: parton shower and later hadronization.
It is well known that even mainstream Monte Carlo programs 
do not  match in this respect sufficiently well the experimental data  for all required
phase space regions \cite{Aad:2016oit}.
This is a complex issue which we can not exhaust\footnote{ 
Large effort is devoted to this aspect of phenomenology thanks to new
data analysis techniques. For example,
new interesting results are obtained thanks to Machine Learning approach
\cite{Larkoski:2017jix,Metodiev:2017vrx}.}.

The discussion of the resulting systematic errors of our method    is out of scope of the 
present paper. It would require evaluation of how 
mismatches of the color and flavour input for parton shower and hadronization
translates into reconstructed jets from simulated detector responses.
This in turn would require the use of 
experimental detector response codes, not available publicly. 
In general,  the {\tt TauSpinner} application domain is  restricted 
to   observables where details of jets,  resulting from partons flavours
or color are not of  importance. This has to be kept in mind.

Let us point that our study examples of the previous sub-sections
are for the cases, where starting and target 
distributions are massively different. In practical applications
we expect {\tt TauSpinner} to be used in configurations where new contributions 
to matrix elements are at the edge of observability. 

If required, it is possible to apply {\tt TauSpinner} in the flavour savvy manner.
Possible solution may follow the method described in Appendix A.
Contributions from distinct flavour configurations can be treated separately 
 only for cases when in experiment 
production files  the flavour configurations are stored, or can be unwinded.

\section{Spin dependent characteristics} \label{sec:spin}
So far we were discussing observables relying on the kinematics  
of final states consisting of four momenta of $\tau$ leptons and 
accompanying two jets. 
Inclusion of   $\tau$ decay products
increase the phase space dimensionality substantially,  making the analysis much more difficult, 
especially  when dependence on selection cuts is taken into account (as  observed  in previous 
sub-sections).

In the following, we will present a few spin dependent results obtained for the $H$ and $X$ samples
within the tight selection cuts. Using {\tt TAUOLA ++ }\cite{Davidson:2010rw} we supplement  these 
samples with $\tau$ decays in the simplest possible mode $\tau^\pm \to \pi^\pm \nu$ with 
 no spin effects included. Spin effects are introduced with the help of {\tt TauSpinner} weights, which are calculated according to the production and decay kinematics (see Refs.~\cite{Czyczula:2012ny,Kaczmarska:2014eoa} 
for the spin weight definition).

\begin{figure}[h!]
 \begin{center}                              
{
 \includegraphics[width=8cm,angle=0]{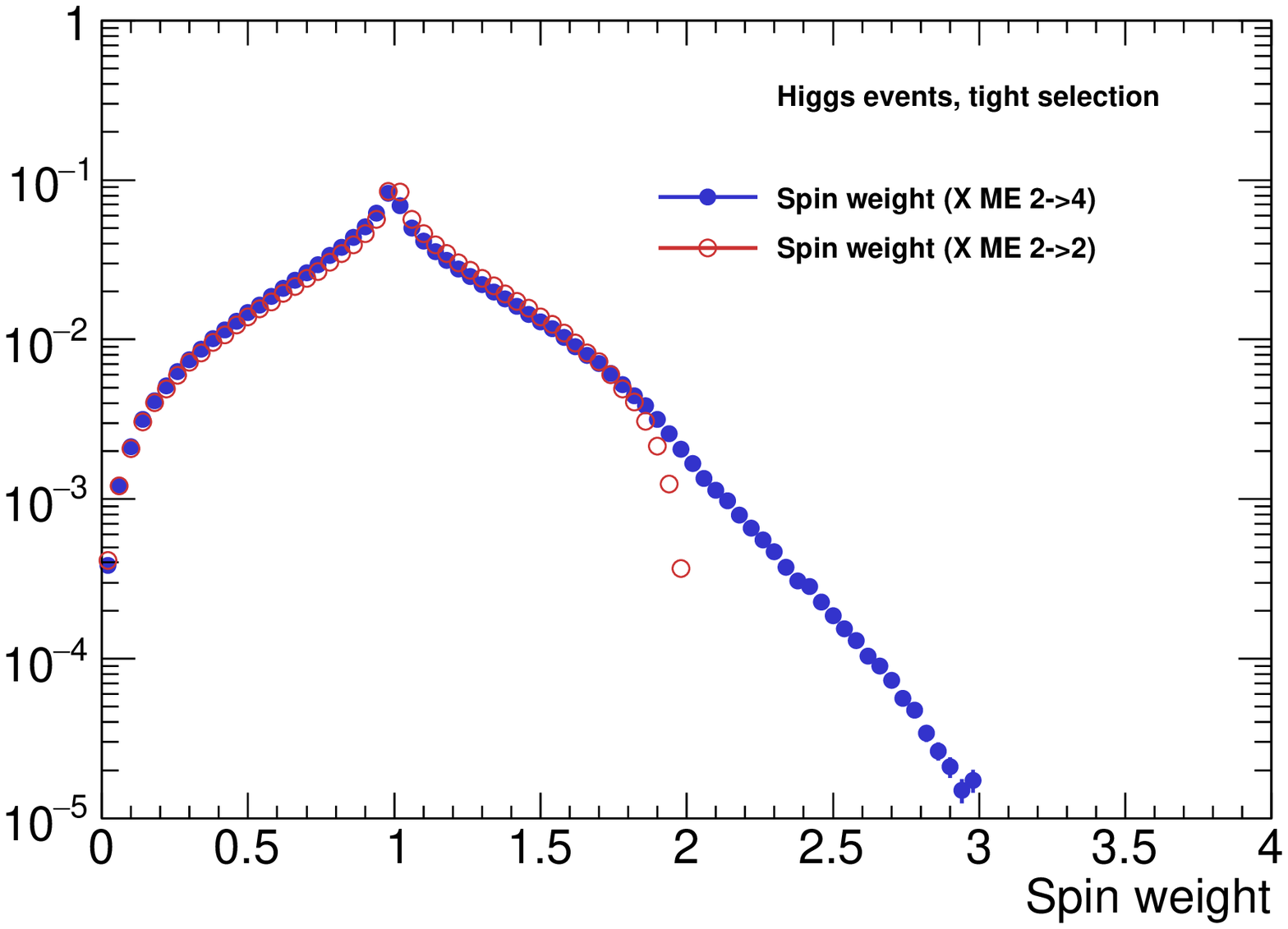}
 \includegraphics[width=8cm,angle=0]{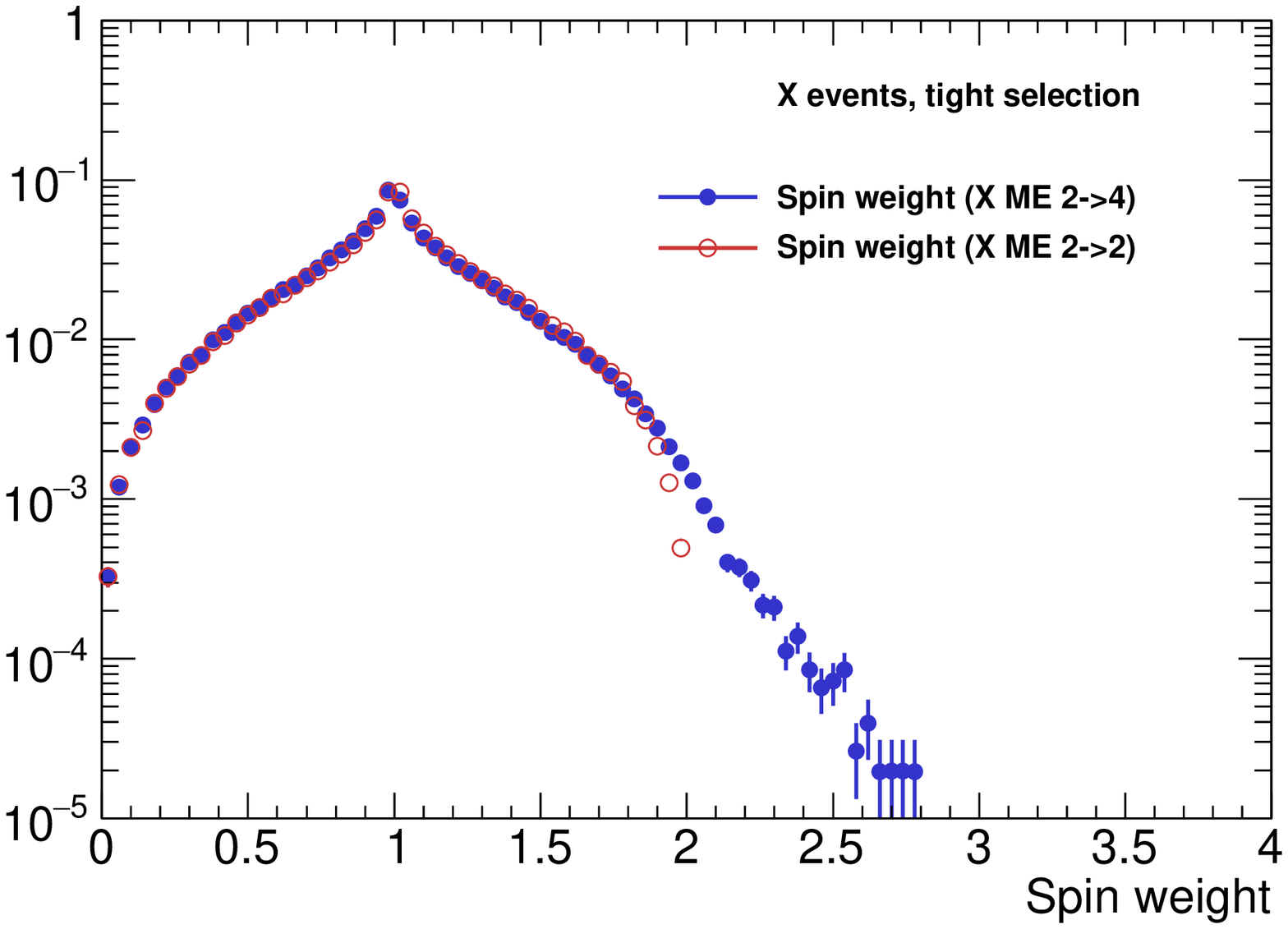}
 }
\end{center}
\caption{Spin weight histograms, normalized to unity, obtained  from $X$ matrix elements 
   for $H$ sample (top plot) and $X$ sample (bottom plot). 
In both cases samples are  
 constrained with tight selection cuts. Red open circles are when  the 
effective Born ($2\to 2$) matrix elements are used and blue full circle points 
are when our new ($2\to 4$) matrix elements are used.
   }\label{fig:spinwt}
\end{figure}

Figure \ref{fig:spinwt} shows  the spin weight histograms for the $H$ and $X$ samples. 
In both cases the spin weights are calculated first using the  matrix element 
for $X$ productions as described in Ref.~\cite{Banerjee:2012ez}, that is 
featuring effective 
Born $2\to 2$ kinematic (open red circles), and  compared with the new calculation in 
which  amplitudes featuring two jet kinematics are taken into account  (blue full circle points)%
\footnote{Here we exploit the virtue of {\tt TauSpinner} which allows, for a given sample, 
to calculate 
weights for different production mechanisms. This feature is of help in validating or rejecting 
theoretical hypotheses. }. 
 In both cases the same $X-\tau\tau$ couplings were used.
As expected (see Eq. 8 from Ref.~\cite{Czyczula:2012ny}), for the  $2\to 2$  case the 
range of spin 
weights is limited to $[0,2]$ since in this process 
there are no couplings which could lead to individual $\tau$ polarization.
In the   $2\to 4$ case  the  spin weight distribution exhibits a tail which  extends beyond $2$ and
covers  most of the allowed $[0,4]$ range. This is due to, e.g., the presence of the  subprocess 
$W^+W^-\to X\to\tau^+\tau^-$  in which $W$'s radiated off  quarks are polarized which has 
impact on $\tau$ polarization.  The tail above $2$,  
although  not so much pronounced,  will manifest itself 
in the distribution of $\tau$ decay products.

Let us now turn to the standard  spin sensitive distribution 
of the ratio $E_\pi/E_\tau$  (a fraction of $\tau$ energy carried by the  decay pion) used in   \cite{Kaczmarska:2014eoa} for benchmarking $\tau$ polarization.
In every case discussed below we will use again 
$X$ production amplitudes to calculate $\tau$ pair density matrix. 
We will do that also for the sample generated with $H$ production amplitudes%
\footnote{Note that the separate treatment of $\tau$ production  from the 
distributions of $\tau$ decay products  enables evaluation of how important
spin effects can be in experimental analyses.}.

\begin{figure}[h!]
 \begin{center}                              
{
 \includegraphics[width=8cm,angle=0]{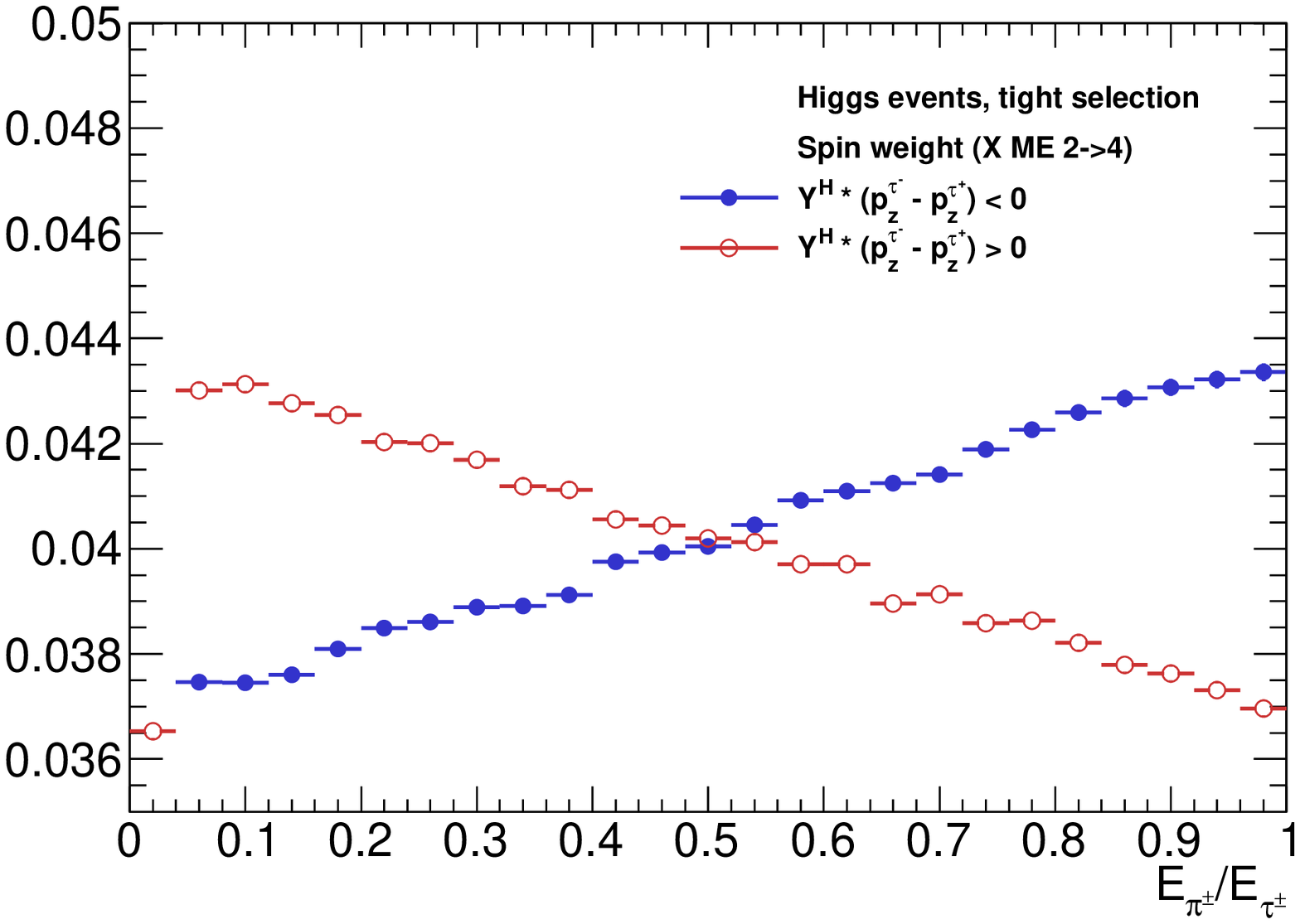}
 \includegraphics[width=8cm,angle=0]{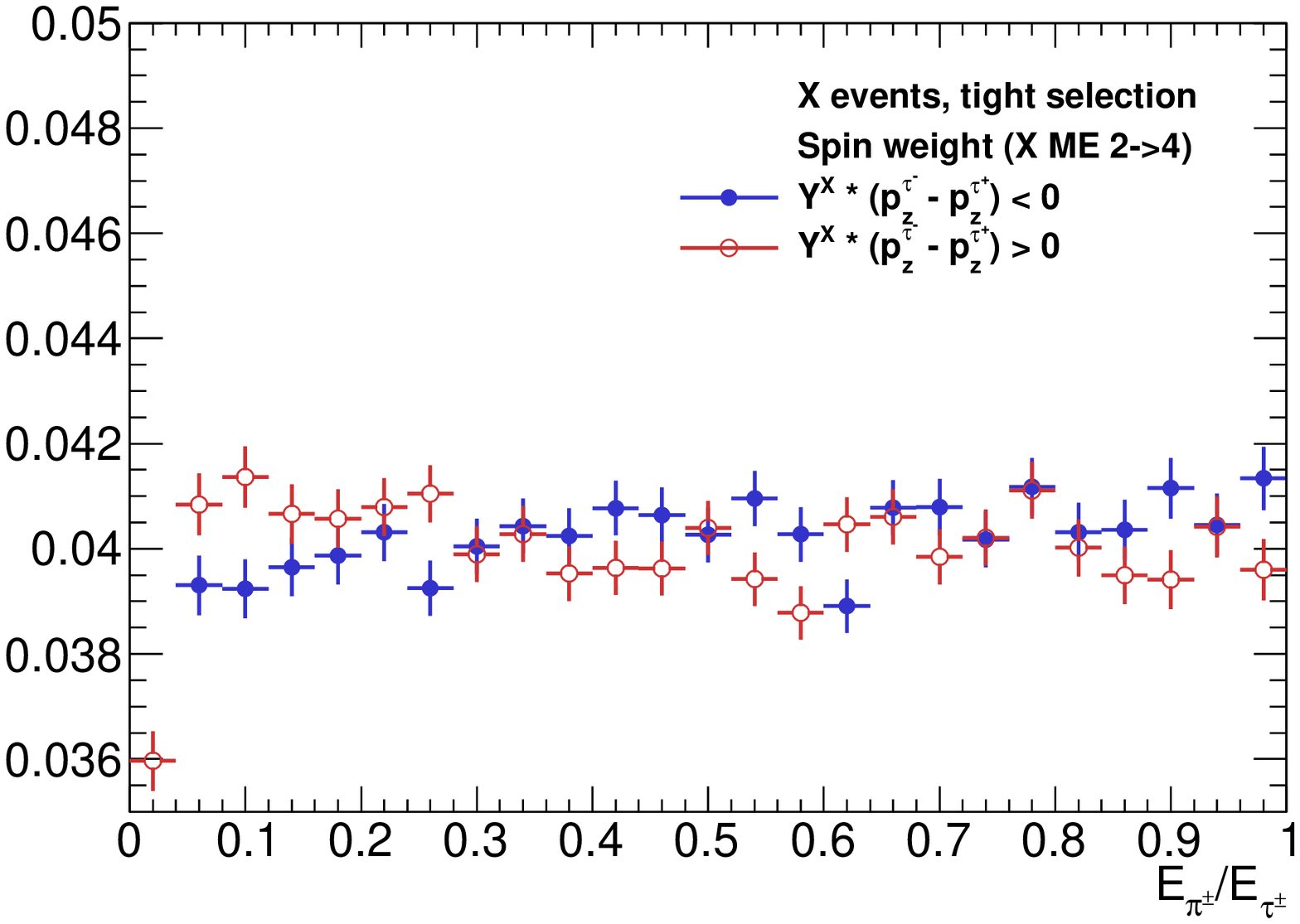}
 }
\end{center}
\caption{Histograms of $E_\pi/E_\tau$ spectra, normalized to unity,  for the $H$ sample (top figure) and for 
the $X$ sample 
(bottom figure).  In all cases $2\to 4$ matrix elements of $X$ exchange are used
to implement spin effects. 
Red open circle points are for additional cut  $Y_X \cdot ( p_Z^{\tau^-}-  p_Z^{\tau^+})>0 $  
and the blue, full circle points for  $Y_X \cdot ( p_Z^{\tau^-}-  p_Z^{\tau^+})<0$. 
Note, that because far less $X$ events survive tight selection, statistical errors
on the bottom plot are larger.}
\label{fig:pispec}
\end{figure}

The $\tau$ polarization can originate from the $X$ production  via VBF process, 
which is asymmetric over the phase-space regions due to the asymmetry of valence $u$ and $d$ quark distributions in the proton. To exhibit the polarization effects we 
have to  sort out events according to the $\tau$ polarization; otherwise the effects will 
average out. Since in the proton there are 
more $u$-type quarks than $d$-type, the $X$ particle produced in the VBF preferentially 
will follow the direction 
of $W^+$ which are right-handed and impart their polarization on $X$ bosons.   
 One can then expect that  $\tau$ 
lepton from $X$ decay will have polarization dependent on $\tau$ direction with 
respect to the $X$ flight direction correlated with its spin polarization.   
Thus it is suggestive to sort events 
according to  positive and negative value of $C=Y_X \cdot ( p_z^{\tau^-}-  p_z^{\tau^+})$, where  
$Y_X$ denotes the $\tau$ lepton pair rapidity and  $p_z^{\tau^-}$, $p_z^{\tau^+}$ are 
the $z$ components of $\tau^\pm$ four-momenta. 
In Fig.~\ref{fig:pispec} events with positive and negative $C$ are plotted separately (the first bin for $C>0$ is  lower  exhibiting the pion mass 
$m_\pi/m_\tau$ effect). 
We observe that spin weights, calculated with the $X$ production amplitude,   
when  applied to the  $H$ sample lead to a larger spin effect, than when applied  to the  
$X$ sample. In the second case the spin effect 
is barely visible\footnote{
It can be used nevertheless to improve exclusion of spin-2 hypothesis.}. 

Our results illustrate the  complexity of multidimensional distributions. 
Even within tight selection 
there is a sizable difference between events of $X$ and $H$ production, which is 
reflected in $\tau$ polarization effects greater for the $H$ sample  
than for $X$ sample, even though the same
$pp \to \tau \tau j j$ matrix elements featuring intermediate $X$ are used in both cases. 

One could argue, that such small spin effect present in Fig. \ref{fig:pispec} for the 
$X$ case is a consequence of substantial contribution from other than VBF channel in 
our samples, thus pointing that
 our cuts may need to be refined. 
However, because of the weight distribution, as  seen in the lower lower plot of 
Fig. \ref{fig:spinwt},  such a refinement is  unlikely to be found  within our tight  
selection, since the tail of events with spin weight exceeding 2 is very small. 
It seems that a better discriminating power between  the   $H\to \tau\tau$ 
and  $X\to \tau\tau$ hypotheses  
can be provided by longitudinal $\tau$-$\tau$ spin correlation, the same as discussed 
already in Refs.~\cite{Banerjee:2012ez,Kaczmarska:2014eoa}.
Nonetheless $\tau$ polarization may offer (minor) help in exclusion of $X$ hypothesis, 
 even in the case when  $X\tau\tau$ couplings are insensitive to parity.

\section{Summary and outlook} \label{sec:summary}
The main  purpose of the paper was to 
 demonstrate   how the new (with respect to the ones used for sample generation) matrix 
 elements  for the 
production of $\tau$-lepton pair accompanied with two jets  in $pp$ collisions
can be used in {\tt TauSpinner} environment to  reweight events. 
 For that purpose, the New Physics matrix element 
for  spin-2 $X$ particle  was implemented as  a user example.

 We have provided numerical tests of 
the algorithm, demonstrating
that starting from the $H$  sample (or $X$ sample), the other one can 
be obtained by applying event-by-event weight calculated from the implemented 
matrix elements. 
We have also addressed possible technical 
difficulties and limitations in implementing the user  code for matrix elements. 
 Even though the {\tt TauSpinner} algorithm in the case of its native and external matrix elements
works similarly, technical aspects due to e.g. rounding errors  and other numerical complications may differ;
thus require individual attention. The  density of events 
to be reweighted may differ from the target one significantly, resulting in a few events with weights massively 
larger than  the ones from other regions of the phase-space.

Limitations  of the algorithm, discussed in sub-section \ref{sec:reliability}, may be observed 
if for example colour or spin configurations 
for original and 
new process play important role for the parton shower. 
Then, reweighting with matrix element of hard process only
 Eq. (\ref{eq:wt_me}), may be too simplistic and factorization 
properties may need to be addressed.
An effort into that direction can be found in Refs.~\cite{Richter-Was:2016mal,Richter-Was:2016avq}.

Let us stress,  that the  {\tt TauSpinner} reweighting can  be repeated 
several times on the same event to obtain multiple  variants of  weights, e.g. due to several
 variants 
of coupling constants, or even completely distinct  $X$ interaction forms.  
In our examples we have used rather 
small  sets of non-zero couplings, see Appendix~\ref{app:troubles},   
in part to simplify  differences in distributions 
of $X$ and $H$ mediated processes. The reweighting algorithm performed better when reduced
region of the phase-space was used for comparisons. 

{  To demonstrate effects sensitive to the $\tau$ lepton polarization
we have chosen $\tau^\pm \to \pi^\pm \nu$ decay mode as a spin analyser. Spin effects 
 originate from the 
$X$ production vertex 
and are embedded in complexity of the multi-body phase-space.   
They turn out to be rather small for our choice of the $X\tau\tau$ couplings.  Nevertheless, they may turn useful in falsifying physics hypotheses   
 alternative to Higgs production and decay processes. }

This paper completes the description of {\tt TauSpinner} functionality, initiated in~\cite{Banerjee:2012ez}
for the   $2 \to 2$ matrix elements
 of New Physics,  now also with 
 the vector-boson-fusion  $2 \to 4$ matrix elements. It  supplements 
examples of {\tt TauSpinner} applications for events with two jets accompanying 
$\tau$-lepton pair production in $pp$ collisions, discussed in~\cite{Kalinowski:2016qcd}.

\vskip 1 cm
\centerline{\bf \Large Acknowledgments}
\vskip 0.5 cm

We thank Tomasz Przedzi\'nski for fruitful discussions and  help with programming aspects of the {\tt TauSpinner} code 
development. 
The work has been supported in part from funds of Polish National Science
Center under decisions UMO-2014/15/B/ST2/00049, by PLGrid Infrastructure of the Academic 
Computer Centre CYFRONET AGH in Krakow, Poland (where  majority of numerical calculations were performed) and by HARMONIA 
project under contract UMO-2015/18/M/ST2/00518 (2016-2019).
MB, JK, ERW and ZW were supported in part by the Research Executive  Agency (REA) of the European Union under the Grant 
Agreement PITNGA2012316704 (HiggsTools). WK was supported in part by the German DFG grant STO 876/4-1. JK thanks the  CERN Theoretical Physics Department for hospitality during the final stage of this work.

\appendix
\section{Example with   $X$ mediated processes: user installation prototype} \label{app:HowToUse}

The purpose of this Appendix is to present how the reweighting with  matrix elements of 
$X$ mediated process, as available 
in the program distribution tar-ball, can be used. It is equally important to 
demonstrate how any other external matrix elements prepared by the user
can be installed:  the $X$ case can serve as a prototype.
The detailed instructions how the reweighting algorithm works and how 
it can be used  for final states of $\tau$ lepton pair and additional two 
jets is given already in 
Appendix A2 of Ref.~\cite{Kalinowski:2016qcd}. In that paper, 
the non Standard Model matrix elements were not discussed.

In the case of our example of spin-2  $X$ mediated matrix elements, 
 the command 
\texttt{
TauSpinner::set\_vbfdistrModif(SPIN2::spin2distr);
}  
is used to   set the pointer to {\tt SPIN2::spin2distr(...,KEY)}. \\
This can be done for the  user own matrix element
routines.
These routines  should over-load the prototype ones for the 
non-standard calculation.

At initialization, when command {\tt  spin2init\_()} is executed the masses and coupling constants for  
\texttt{SPIN2::spin2distr(...,KEY)} 
calculation 
are set.  Later, for every event the algorithm  makes the choice for the actual 
matrix elements used in the weight calculation: 
it  evaluates and pass to the user provided function its internal parameter {\tt KEY}. 
The {\tt KEY = 0, 2} corresponds to  Drell-Yan like processes of the Standard Model and anomalous (user provided) 
matrix element.
Analogously {\tt KEY = 1, 3} is for the Higgs of the Standard Model and
(user provided) matrix element  for anomalous narrow resonance.  
 The code  will choose  between 
Higgs  and Drell-Yan background amplitudes on the basis of PDG identifier  of  the intermediate resonance found 
in the event record\footnote{
Let us stress that {\tt TauSpinner} assumes that the events sample used as an input is of the
Standard Model type. It determines for every event, if it is of the Higgs type, 
by checking if intermediate state  {\tt PDGid=25}.
Otherwise, default Drell-Yan production will be assumed.
}. 
For {\tt X.pdgID} $\ne 25$ it will set {\tt KEY=0} for Standard Model 
(A) - that is for denominator of $wt_{prod}^{A \rightarrow B}$ given in Eq.(\ref{eq:wt_me}) and {\tt KEY=2} for (B), weight numerator.
For {\tt X.pdgID} $= 25$ it will set {\tt KEY=1} for (A)
and {\tt KEY=3} for (B).
This is why  user-provided function for the matrix element calculation must have a {\tt KEY} parameter among its arguments.

 The interface assumes that the event sample is for the SM i.e. as of type (A) used in weight denominator while the user provided function is of type (B), and accordingly the weight $wt_{prod}^{A \rightarrow B}$  is calculated.
If the analyzed event  sample is (as for  Fig.~\ref{fig:five}) of type (B), then inverse of 
the weight given by Eq. (\ref{eq:wt_me}) and calculated by {\tt TauSpinner} should be used for reweighting.

So far, we have not discussed spin polarization and spin correlations between outgoing $\tau$ leptons. 
If a generated sample features spin correlations and  polarizations of 
$\tau$-leptons, then it has to be taken into account. The event weight 
needs to be supplemented with spin weights of the Standard Model and New Physics
 calculation
\begin{equation}
wt_{prod}^{A \rightarrow B} \to wt_{prod}^{A \rightarrow B}\; \cdot \; {\tt WT/WT0},
\label{eq:wt0}
\end{equation}
where {\tt WT} and  {\tt WT0} 
 denote usual spin weight as of Eq. (5) from Ref.~\cite{Czyczula:2012ny} calculated  
first for (B) and then for (A). They
can be obtained as shown in  the  extract from the code of the demonstration program.

\subsection{Example of the demonstration  program  for $X$ mediated processes}
\label{sect:MainProgram}
The detailed instruction for the demonstration program is given in 
Appendix A4 of Ref.~\cite{Kalinowski:2016qcd}. However, some changes were 
required to allow the use of
 $X$ mediated amplitudes and corresponding reweighting. 
The distribution tar.ball \cite{tauolaC++} starting from version  of Jul. 11 2017, 
includes necessary provisions, {  which are commented out.
For $X$ mediated processes they have to be activated in the following  files of  the directory
{\tt TAUOLA/TauSpinner/examples/example-VBF}:}
\begin{itemize}
\item
the  user example program {\tt  example-VBF.cxx},
\item
the  prototype {\tt method} {\tt read\_particles\_for\_VBF.cxx } to read in events stored in {\tt 
HepMC} format,  prepared specifically for {\tt MadGraph5}
generated events,
\item the {\tt SPIN2} directory which contains matrix elements,
\item the {\tt README} file in the {\tt TAUOLA\slash{}TauSpinner\slash{}examples\slash{}example-VBF/SPIN2}, which contains all information for using the 
$X$ matrix elements implementation which is as follow:\\
 \phantom{MM} $-$ open {\tt Makefile} in {\tt SPIN2} directory and follow instructions there,\\
 \phantom{MM} $-$  introduce  a link to the SPIN2 library into configuration of {\tt TAUOLA/TauSpinner/examples}
\end{itemize}
To activate SPIN2 case in {\tt example-VBF.cxx}: 
\begin{itemize}
\item modify {\tt TAUOLA/TauSpinner/examples/Makefile}:  uncomment lines containing string SPIN2,
\item modify {\tt TAUOLA/TauSpinner/examples/example-VBF\slash{}example-VBF.cxx}:  again look for 
   commented lines which contain string spin2 or SPIN2 and uncomment them.
   Note that {\tt example-VBF.cxx} will use {\tt method}  {\tt TauSpinner::set\_vbfdistrModif(SPIN2::spin2distr)},
   but call on the {\tt F77} {\tt spin2init\_()} routine for initialization of SPIN2 library constants is necessary.
 \item  path to  SPIN2 library has to be exported:\\
  {\footnotesize export {\tt LD\_LIBRARY\_PATH="/home...TAUOLA\slash{}TauSpinner\slash{}examples\slash{}example-VBF\slash{}SPIN2\slash{}lib:\$LD\_LIBRARY\_PATH"} }
\item {\tt Makefile} of {\tt SPIN2} directory has to be executed,
\item the examples of {\tt TAUOLA/TauSpinner/examples} and
                       {\tt TAUOLA/TauSpinner/examples/example-VBF} 
                       with SPIN2 library activated are ready to be  compiled and linked.
\end{itemize}
 
For the user own external matrix elements, the scheme as of
  SPIN2 should be treated as a prototype which need to be followed. 
The  changes into the demonstration program {\tt  example-VBF.cxx} explained above provide  necessary instructions.
For further technical details we provide below  an extract from the code.


\subsection{Extract from the code  {\tt example-VBF.cxx}. }

An extract from the   {\tt TAUOLA/TauSpinner/examples\slash{}example-VBF/example-VBF.cxx} file of the
distribution tar-ball.
 Less important, at the  first reading parts, are dropped.
On the other hand, code for calculation of spin weight contribution (Eq.~\ref{eq:wt0}) is listed.

\onecolumn
{\scriptsize
\begin{verbatim}


//-----------------------------------------------------------------

// For SPIN2 code placed in directory example-VBF:
// #include "spin2distr.h"  // will work once `export SPIN2LIB=...' 

---------------------------------------------------------------------
/** Example function that can be used to modify/replace matrix element calculation of vbfdist present in SPIN2/ME/spin2distr.cxx */
//  double spin2distr(int I1, int I2, int I3, int I4, int H1, int H2, double P[6][4], int KEY, double vbfdistr_result)

int main(int argc, char **argv) {

    if(argc<2) {
        cout<<"Usage:    "<<argv[0]<<" <input_file> [<events_limit>]" << endl;
        cout<<"Consider: "<<argv[0]<<" events-VBF.dat 10" << endl;
        exit(-1);
    }

    char *input_filename = argv[1];
    int   events_limit   = 0;
    if(argc>2) events_limit = atoi(argv[2]);

    //---------------------------------------------------------------------------
    //- Initialization ----------------------------------------------------------
    //---------------------------------------------------------------------------

    // Initialize Tauola
    Tauola::initialize();
    Tauola::spin_correlation.setAll(false);

    // Initialize random numbers:
    // ##1##
    // Important when you re-decay taus: set seed fortauola-fortran random number generator RANMAR
    // int ijklin=..., int ntotin=..., int ntot2n=...; /
    // Tauola::setSeed(ijklin,ntotin,ntot2n);
    // Tauola::setSeed(time(NULL), 0, 0);

    // ##2##
    // Important when you use attributed by TauSpinner  helicities
    // Replace C++ Tauola Random generator with your own (take care of seeds). Prepared method: 
    // gen.SetSeed(time(NULL));
    // Tauola::setRandomGenerator( randomik );

    // Initialize LHAPDF
    // string name="MSTW2008nnlo90cl.LHgrid";
    string name="cteq6ll.LHpdf";
    // choice used for events-VBF.lhe bring tiny variation, thus it is not 
    // string name="MSTW2008nlo68cl.LHgrid"; //             statistically important
    LHAPDF::initPDFSetByName(name);
\end{verbatim}
{ 
\begin{verbatim}
    // CMSENE is the center of mass energy  used in PDF calculations; only if Ipp = true
    // and only for Z/gamma*
    //  Ipp - true for pp collision; otherwise polarization
    //  of individual taus from Z/gamma* is set to 0.0

    // Ipol - relevant for Z/gamma* decays 
    //  0 - events generated without spin effects                
    //  1 - events generated with all spin effects               
    //  2 - events generated with spin correlations and <pol>=0  
    //  3 - events generated with spin correlations and
    // polarization but missing angular dependence of <pol>

  
   // nonSM2  option- 1/0 extra term in cross section, density matrix on/off   
   // nonSMN  option- 1/0 extra term in cross section, for shapes only on/off 
\end{verbatim}
   // For definition of {\tt EW and QCD schemes} see Ref.~\cite{Kalinowski:2016qcd}. 
  
}    
\begin{verbatim}
    double CMSENE = 13000.0;  // 14000.0;
    bool   Ipp    = true;
    int    Ipol   = 1;
    int    nonSM2 = 0;
    int    nonSMN = 0;

    // Initialize TauSpinner
    initialize_spinner(Ipp, Ipol, nonSM2, nonSMN,  CMSENE);

    int ref=1;      // EW scheme to be used for default vbf calculation. 
    int variant =1; // EW scheme to be used in optional matrix element reweighting (nonSM2=1). Then
                    // for vbf calculation, declared above prototype method vbfdistrModif 
                    // (or user function)will be used. 
                    // At its disposal result of calculation with variant of  EW scheme will be available.
    vbfinit_(&ref,&variant);

    int QCDdefault=1; // QCD scheme to be used for default vbf calculation.
    int QCDvariant=1; // QCD scheme to be used in optional matrix element reweighting (nonSM2=1).
    setPDFOpt(QCDdefault,QCDvariant);

    // Set function that modifies/replaces Matrix Element calculation of vbfdistr
    // TauSpinner::set_vbfdistrModif(vbfdistrModif);

    // Set function that modifies/replaces Matrix Element calculation of vbfdistr with code of SPIN2
    // spin2init_(&ref,&variant);
    // TauSpinner::set_vbfdistrModif(SPIN2::spin2distr);
 
    // Set function that modifies/replaces alpha_s calculation of vbfdistr
    // TauSpinner::set_alphasModif(alphasModif);

    // Open I/O files  (in our example events are taken from "events.dat")
    HepMC::IO_GenEvent input_file(input_filename,std::ios::in);

    if(input_file.rdstate()) {
        cout<<endl<<"ERROR: file "<<input_filename<<" not found."<<endl<<endl;
        exit(-1);
    }
    int events_read =0;

    int    events_count = 0;
    double wt_sum       = 0.0;

    //---------------------------------------------------------------------------
    //- Event loop --------------------------------------------------------------
    //---------------------------------------------------------------------------
    while( !input_file.rdstate() ) {
        double    WT      = 1.0;
        double    W[2][2] = { { 0.0 } };

        SimpleParticle p1, p2, X, p3, p4, tau1, tau2;
        vector<SimpleParticle> tau1_daughters, tau2_daughters;

        int status = read_particles_for_VBF(input_file,p1,p2,X,p3,p4,tau1,tau2,tau1_daughters,tau2_daughters);
        ++events_read;

    // WARNING: meaning of status may depend on the variant of  read_particles_for_VBF().
        if( status == 1 ) break;                      // variant A
        //        if     ( status == 0 ) break;       // variant B
        //        else if( status == 1 ) { continue;} // variant B


        // setNonSMkey(0);  // to calculate spin weight of Standard Model
        // double WT0 = calculateWeightFromParticlesVBF(p3, p4, X, tau1, tau2, tau1_daughters, tau2_daughters);

        // setNonSMkey(1);  // e.g. for use of SPIN2 ME and calculate all weights
        WT = calculateWeightFromParticlesVBF(p3, p4, X, tau1, tau2, tau1_daughters, tau2_daughters);

        // double WTME=getWtNonSM();        // matrix element weight
        //        WTME=WTME*WT/WT0;         // factor to take into account spin correlations of tau-tau pair decays
        wt_sum += WT;
        ++events_count;
        if( events_count%100 == 0 ) cout << "EVT: " << events_count << endl;
        if( events_limit && events_count >= events_limit ) break;
    }
    
    cout<<endl<<"No of events read from the file: "<<events_read<<endl;
    cout<<endl<<"No of events processed for spin weight: "<<events_count<<endl;
    cout<<      "WT average for these processed events: "<<wt_sum/events_count<<endl;
\end{verbatim}
}
\FloatBarrier
\twocolumn

\section{ Details of {\tt MadGraph5}  initialization. }\label{app:troubles}
The settings and parameters  for the  {\tt MadGraph5} \cite{Alwall:2014hca}
generation of $H$ and spin-2 $X$ event samples used in Section~\ref{sec:numerical}
are collected in the web page \cite{WebPageSoinner2j}.
We remind the reader that these two generation parameter sets must be carefully matched, with the corresponding ones 
in the initialization of native {\tt TauSpinner}  (file {\tt ../TauSpinner/src/VBF\slash{}VBF\_init.f}) and external matrix elements. 
In our example, it is the file {\tt ../TauSpinner/examples\slash{}example-VBF\slash{}SPIN2\slash{}ME/SPIN2\_init.f }. 
A similar input consistency check will have to be performed whenever user own New Physics matrix elements
are installed.

Let us recall some details how the code for matrix elements of  $X$ mediated processes were prepared. 
After the default (automatic) initialization with  \texttt{MadGraph5} settings
{\small
\begin{verbatim}
set group_subprocesses Auto
set ignore_six_quark_processes False
set loop_optimized_output True
set low_mem_multicore_nlo_generation False
set loop_color_flows False
set gauge unitary
set complex_mass_scheme False
set max_npoint_for_channel 0
import model sm
define p = g u c d s u~ c~ d~ s~
define j = g u c d s u~ c~ d~ s~
define l+ = e+ mu+
define l- = e- mu-
define vl = ve vm vt
define vl~ = ve~ vm~ vt~
\end{verbatim}
}
\noindent we produced the event generation code with the  sets of commands
given in Table \ref{tab:mad}.
\begin{table*}[h]
\begin{center}
\begin{tabular}{>{\ttfamily}{l}>{\ttfamily}{l}} 
    \toprule
    \multicolumn{1}{c}{Higgs}       & \multicolumn{1}{c}{X}\\
    \midrule
    import model sm-ckm & import model spin2\_w\_CKM\_UFO \\
    generate p p > j j h  QED=99,                    & 
    generate p p > j j x QCD=0 QED=2 NPVV=1                          \\
         \hspace{1cm}  h > ta+ ta-                         & \hspace{1cm} NPll=1,  x > ta+ ta-                                     \\
    output h\_dir                      & output spin2\_dir                \\
    \bottomrule
\end{tabular}
\end{center}
\caption{Commands to generate with the help of \texttt{MadGraph5} the
source code for $H$ and $X$ amplitudes calculation. \label{tab:mad}}
\end{table*}

\noindent 
The corresponding spin-2 \texttt{param\_card.dat} file contains an additional b
lock called \texttt{spin2} for model specific couplings%
\footnote{
 Of the couplings of X to gauge bosons only the CP-even ones were
implemented in the SPIN2 library. They are denoted in the code with a with
suffix \texttt{Even} and correspond to couplings $g_{Xii}$, $i=B,W,g$, in Eq.~(\ref{eq:lag_gauge_eigenstates}).
}.
{\small
\begin{verbatim}
Block spin2 
    4 1.000000e+00 # gXtautauM 
    5 1.000000e+00 # gXtautauP 
    6 0.000000e+00 # gXqqM 
    7 0.000000e+00 # gXqqP 
    8 0.000000e+00 # gXggEven 
    9 0.000000e+00 # gXggOdd 
   10 1.000000e+00 # gXWWEven 
   11 0.000000e+00 # gXWWOdd 
   12 0.000000e+00 # gXBBEven 
   13 0.000000e+00 # gXBBOdd 
\end{verbatim}
}
\noindent as well as a block containing  information about the quantum numbers~ \cite{Alwall:2007mw}
 of the spin 2 object.
{\small
\begin{verbatim}
Block QNUMBERS 5000002  # x 
  1 0  # 3 times electric charge
  2 5  # number of spin states (2S+1)
  3 1  # colour rep (1: singlet, 3: triplet, 8: octet)
  4 0  # Particle/Antiparticle distinction (0=own anti)
\end{verbatim}
}
\noindent In addition, the block containing Wolfenstein parametrization of 
the CKM matrix is called \texttt{ckmblock} in the Spin-2 
\texttt{param\_card.dat} (as opposed to name \texttt{wolfenstein} in the SM one).

The \texttt{run\_card.dat} files used subsequently for generation of both $H$ and $X$ events 
are available from \cite{WebPageSoinner2j}.

\
\subsection{Setting precision parameter of {\tt MadGraph} initialization.  }

The weights calculated with external  new matrix elements may produce a tail of rare, 
 high weight events. This may  lead to   excessively large statistical errors, sometimes  poorly monitored in the histograms. {  Their origin is from 
sparsely populated phase space regions in which 
some events receive high weight due to
some resonance/collinear configuration of the new process,
see. e.g.   discussion of Figs. 3 and 4 in Section 4.2 
of Ref.~\cite{Kalinowski:2016qcd}.  

Let us now comment on} seemingly similar observations but of a fundamentally different origin.
\begin{figure}[t!]
 \begin{center}                               
{
  \includegraphics[width=8.cm,angle=0]{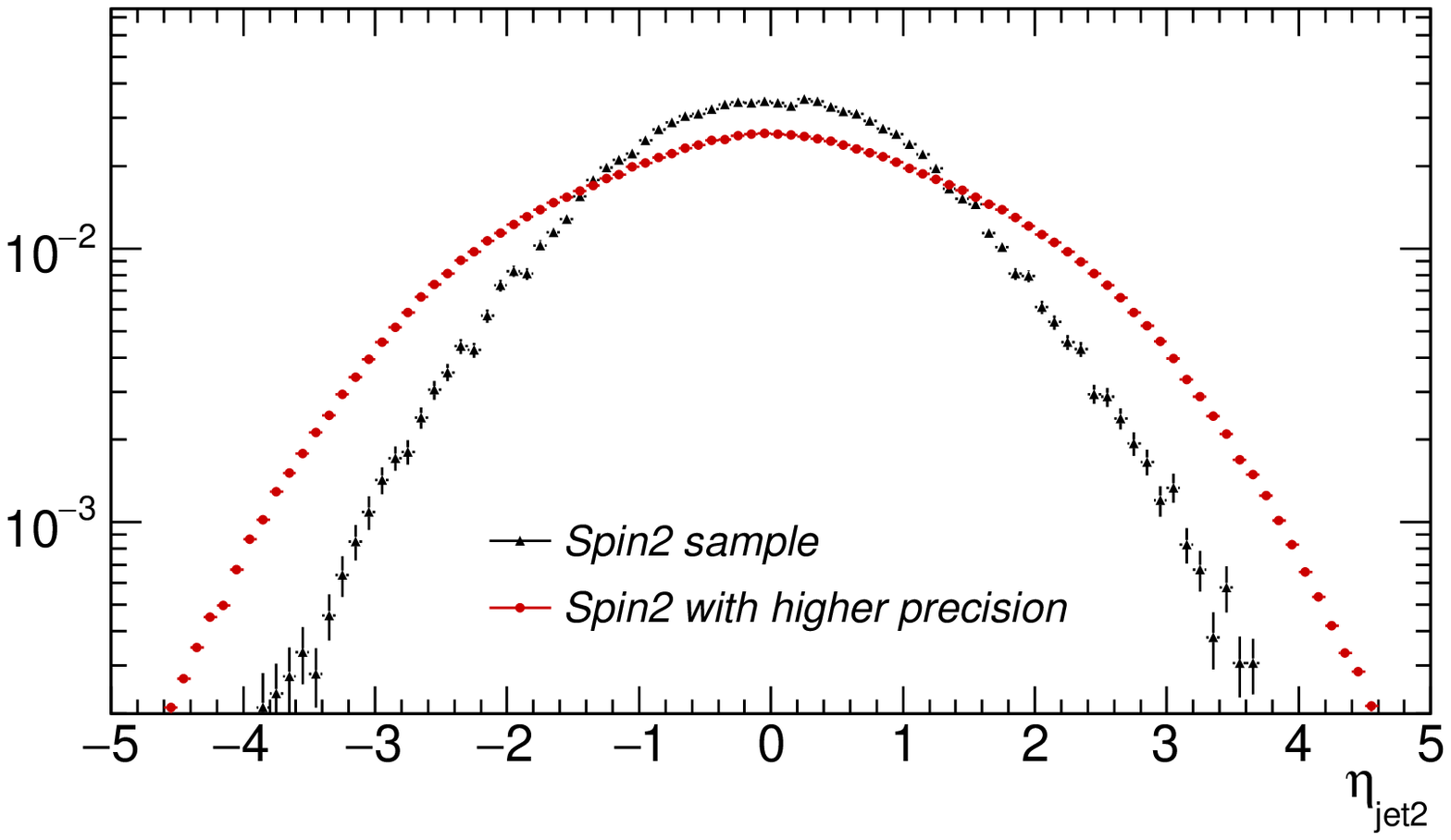}
   \includegraphics[width=8.cm,angle=0]{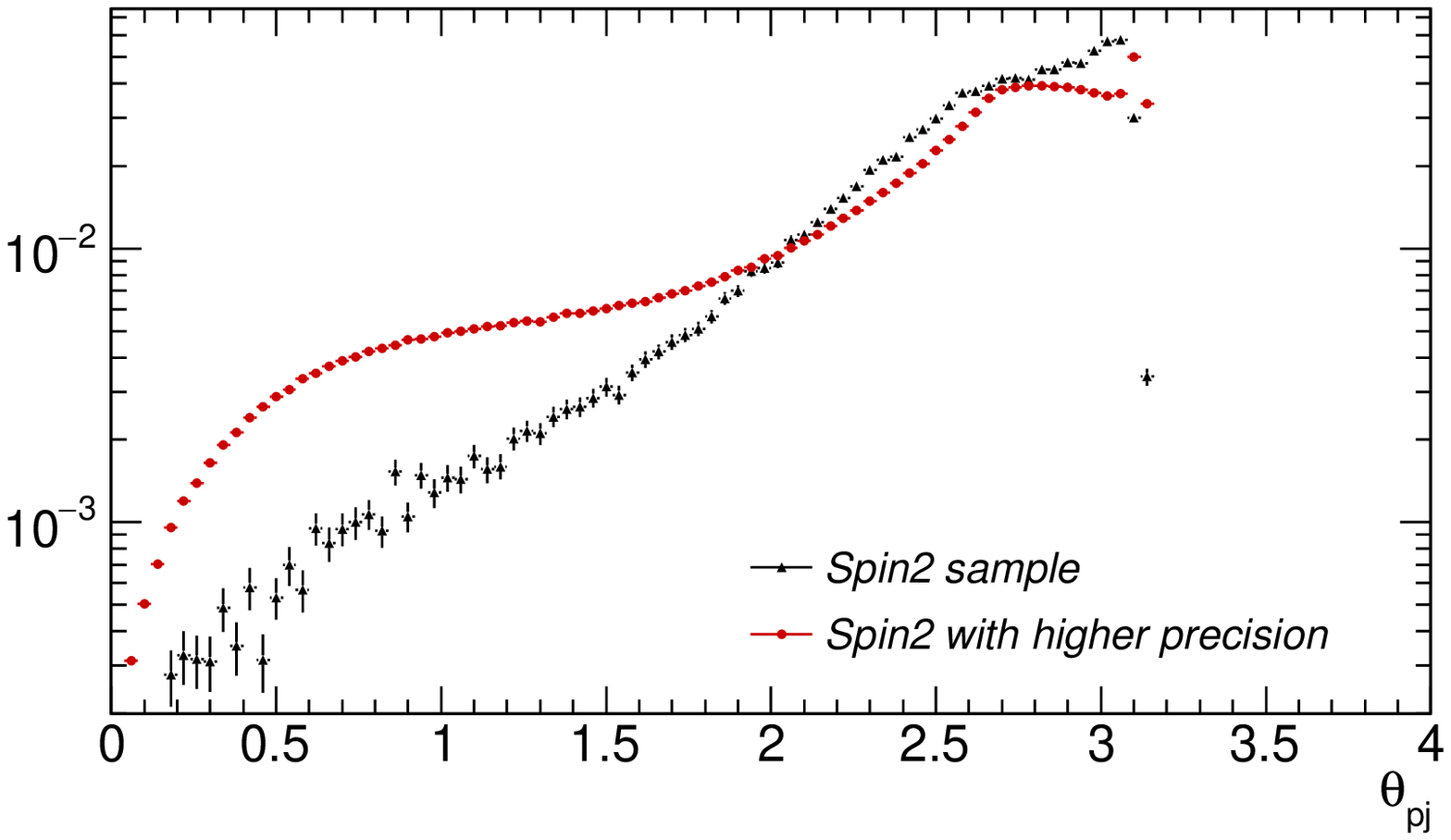}
}
\end{center}
\caption{Histograms of  $\eta_{jet2}$ and $\theta_{pj}$ distributions, normalized to unity,  
from   {\tt MadGraph} generated $X$ samples. Black points are for the 
default 8 iterations used at initialization, and 
red (grey) points for  99 
iterations.  }\label{fig:accu}
\end{figure}
In the early steps of tests, for reweighting of $X$ to $H$ samples,
 we have encountered a mismatch between reweighted and reference distributions  generated in the gridpack mode. 
It turned out that the 
problem was related to the accuracy of samples  generated 
by {\tt Madgraph5} based on {\tt MadEvent}~\cite{Maltoni:2003} algorithm. 
At {\tt MadEvent} core lies the diagram enhancement method
which separates the integration into a sum of integrals whose singularity structure is
dictated by a single Feynman diagram topology. Each of these individual integrals, referred
to as integration channel, 
is then integrated using an appropriate phase-space parametrisation for its underlying 
structure.  
A {\tt MadEvent}
run involves two steps. 
The first one is referred to as the survey
and consists of computing  cross-sections for each integration channel down to a given accuracy.
{Calling the \texttt{generate$_-$events} script invokes a series of four commands: \texttt{survey}, \texttt{combine$_-$events}, \texttt{store$_-$events}, \texttt{create$_-$gridpack}, in which by default \texttt{survey} should reach precision of 0.01  }
 within 8 iterations, where the first iteration consists of 
2000 points. 
However, with this default iteration number the required accuracy of 0.01 was not achieved and the kinematical distributions for $X$ samples did not show 
a good  behavior. { 
After invoking the four commands one by one with the number of iterations  in \texttt{survey}\footnote{The syntax of setting \texttt{survey} options can be found by invoking the \texttt{help survey} command in the \texttt{madevent} shell.}
  increased to 99,  }
the reweighted distributions were 
 matching the
reference ones (as is the case of Figs. \ref{fig:four} and \ref{fig:five}). One can see 
in  Fig.~\ref{fig:accu}, that the requirement for precision for chosen distributions $\eta_{jet2}$ and $\theta_{pj}$ 
 turned out to be particularly demanding. 

\FloatBarrier
\providecommand{\href}[2]{#2}\begingroup\raggedright\endgroup

\end{document}